\documentclass[sigconf]{acmart}

\usepackage{amsmath}
\usepackage{booktabs}
\usepackage{multirow}
\usepackage{multicol}
\usepackage[normalem]{ulem}
\useunder{\uline}{\ul}{}
\usepackage{appendix}
\usepackage{balance}
\usepackage{makecell}
\usepackage{color}

\usepackage[linesnumbered,ruled,vlined, noend]{algorithm2e}

\usepackage{enumitem}
\setitemize[1]{topsep=0pt}

\newcommand{\stitle}[1]{\vspace{1ex} \noindent{\bf #1}}

\AtBeginDocument{%
  \providecommand\BibTeX{{%
    \normalfont B\kern-0.5em{\scshape i\kern-0.25em b}\kern-0.8em\TeX}}}

\copyrightyear{2023}
\acmYear{2023}
\setcopyright{acmlicensed}
\acmConference[KDD '23]{Proceedings of the 29th ACM SIGKDD Conference on Knowledge Discovery and Data Mining}{August 6--10, 2023}{Long Beach, CA, USA}
\acmBooktitle{Proceedings of the 29th ACM SIGKDD Conference on Knowledge Discovery and Data Mining (KDD '23), August 6--10, 2023, Long Beach, CA, USA}
\acmPrice{15.00}
\acmDOI{10.1145/3580305.3599324}
\acmISBN{979-8-4007-0103-0/23/08}

\begin{document}

\title{Efficient Bi-Level Optimization for Recommendation Denoising}

\author{Zongwei Wang}
\email{zongwei@cqu.edu.cn}
\orcid{0000-0002-9774-4596}
\affiliation{%
  \institution{Chongqing University}
  \country{China}
}

\author{Min Gao}
\email{gaomin@cqu.edu.cn}
\authornote{Corresponding author}
\affiliation{%
  \institution{Chongqing University}
  \country{China}}

\author{Wentao Li*}
\email{wentaoli@hkust-gz.edu.cn}
\affiliation{%
  \institution{The Hong Kong University of Science and Technology (Guangzhou)}
  \country{China}}

\author{Junliang Yu}
\email{jl.yu@uq.edu.au}
\affiliation{%
  \institution{The University of Queensland}
  \country{Australia}}

\author{Linxin Guo}
\email{guolinxin@cqu.edu.cn}
\affiliation{%
  \institution{Chongqing University}
  \country{China}
}

\author{Hongzhi Yin}
\email{h.yin1@uq.edu.au}
\affiliation{%
  \institution{The University of Queensland}
  \country{Australia}}








\begin{abstract}
The acquisition of explicit user feedback (e.g., ratings) in real-world recommender systems is often hindered by the need for active user involvement. To mitigate this issue, implicit feedback (e.g., clicks) generated during user browsing is exploited as a viable substitute. However, implicit feedback possesses a high degree of noise, which significantly undermines recommendation quality. While many methods have been proposed to address this issue by assigning varying weights to implicit feedback, two shortcomings persist: (1) the weight calculation in these methods is iteration-independent, without considering the influence of weights in previous iterations, and (2) the weight calculation often relies on prior knowledge, which may not always be readily available or universally applicable.

To overcome these two limitations, we model recommendation denoising as a bi-level optimization problem. The inner optimization aims to derive an effective model for the recommendation, as well as guiding the weight determination, thereby eliminating the need for prior knowledge. The outer optimization leverages gradients of the inner optimization and adjusts the weights in a manner considering the impact of previous weights. To efficiently solve this bi-level optimization problem, we employ a weight generator to avoid the storage of weights and a one-step gradient-matching-based loss to significantly reduce computational time. The experimental results on three benchmark datasets demonstrate that our proposed approach outperforms both state-of-the-art general and denoising recommendation models. The code is available at https://github.com/CoderWZW/BOD.
\end{abstract}







\maketitle

\section{INTRODUCTION}
Recommender systems~\cite{56wang2019enhancing,55guo2019streaming, 48lu2018between,57yin2019social}, which leverage historical behavioral data of users to discover their latent interests, have been highly successful in domains such as E-commerce~\cite{02yu2021self}, for improving user experience and driving incremental revenue.
While the explicit user feedback (e.g., ratings) is the best fuel for recommender systems, its acquisition is often impeded by the need for active user participation. Hence, implicit feedback (e.g.,
clicks) generated during user browsing is exploited as a viable
substitute~\cite{12DBLP:conf/sigir/Gao0HCZFZ22,24wang2018minimax}.
For implicit feedback, an observed user-item interaction is generally regarded as a \textit{positive sample}, while an unobserved interaction is deemed as a \textit{negative sample}~\cite{25yu2021socially}. However, implicit feedback is plagued by a significant level of noise~\cite{30wang2021implicit,32wu2021ready,47chen2021autodebias} as evidenced by the following factors:
(1) users may inadvertently interact with certain items, giving rise to \textit{false positive samples}~\cite{27bian2021denoising,28pan2013gbpr} due to curiosity or misclicks; (2) users may encounter situations where they lack exposure to an item, leading to \textit{false negative samples} ~\cite{12DBLP:conf/sigir/Gao0HCZFZ22}, despite having a positive preference for it. The presence of such noise exacerbates the overestimation and underestimation of some certain user preferences. Hence, it is imperative to mitigate this noise in order to enhance the accuracy of recommendations~\cite{30wang2021implicit}.

A prevalent approach to mitigating the impact of noise in implicit feedback involves utilizing auxiliary information such as attribute features ~\cite{44zhang2022neuro} and external data (e.g., social networks)~\cite{02yu2021self}. 
However, the acquisition of such information may be hindered by privacy concerns~\cite{11wang2021denoising}.
In the absence of auxiliary information, current denoising methods often employ a weighting strategy in which the samples are assigned varying weights iteratively~\cite{11wang2021denoising}. Specifically, this is achieved through the repetitive execution of two phases: first training the recommendation model using the initial weight assignments, and then recomputing the sample weights based on the output of the recommendation model. The weights reflect the level of noise in each sample, as well as their contribution to the recommendation~\cite{12DBLP:conf/sigir/Gao0HCZFZ22}, thus enabling effective denoising.

Despite the demonstrated efficacy, current research still faces two limitations: 
(1) the weight assignment strategy is performed in an iteration-independent manner ~\cite{11wang2021denoising}, leading to the random re-initialization of weights. This implies that the previously computed weights, which possess the ability to express the confidence level regarding the presence of noise in the samples, are disregarded during the calculation of sample weights in the current iteration. As a result, the initialization weights in every iteration is mere blind guesses, while the valuable information generated during previous iterations is not utilized to its full potential; (2) existing methods often resort to \textit{prior knowledge} for determining sample weights. For example, Wang \textit{et al.}~\cite{11wang2021denoising} claim that a sample with high loss is more likely to be noisy, whereas~\cite{10wang2022learning} argues that a noisy sample varies greatly in loss across multiple recommendation models. Nevertheless, prior knowledge may not be readily available, and its applicability may vary in different situations.

\begin{figure*}[h]
	\centering
	\includegraphics[width=0.75\linewidth]{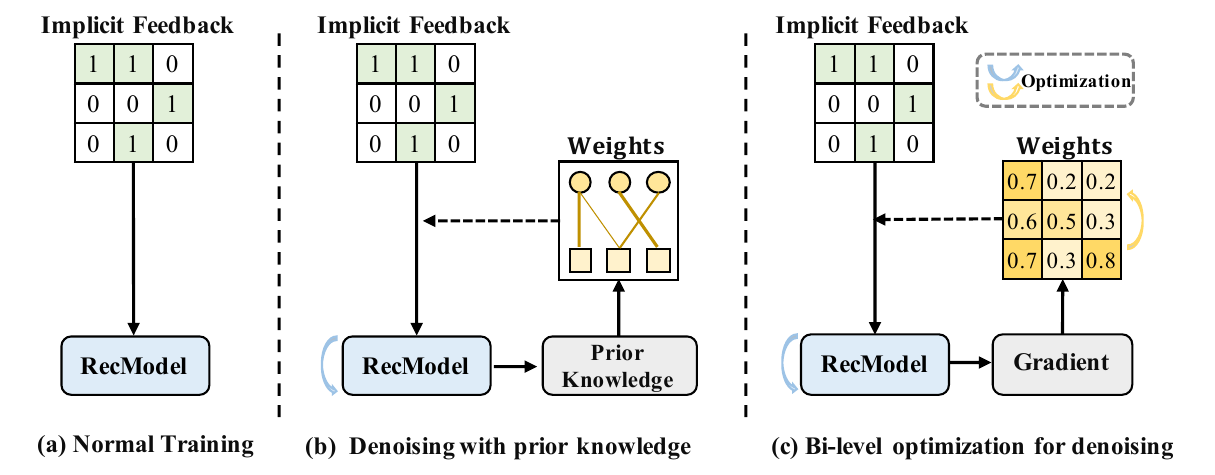}
	\caption{The comparison among (a) Normal recommendation model training without denoising, (b) Denoising training using prior knowledge, and (c) Bi-level optimization for denoising. (b) and (c) are for recommendation denoising without extra information, where the weights in (b) are initialized for recommendation model training in each iteration, but the weights in (c) are stored as a global variable set.}
	\label{introduction}
\end{figure*}

To address the limitations inherent in current methods, we model recommendation denoising as a bi-level optimization problem, which consists of two interdependent optimization sub-tasks: an inner optimization for deriving an effective recommendation model and an outer optimization for learning the weights while considering the impact of previous weights. The logic behind the bi-level optimization can be succinctly explained by two considerations. Firstly, by retaining the weights learned by the outer optimization as global variables, previous adjustments to the weights can be persistent, enabling the sharing of weight information throughout model training. Secondly, it inherently unveils the fundamental goal of recommendation denoising: augmenting recommendation accuracy to guide the denoising procedure. The inner optimization phase primarily focuses on enhancing the accuracy of the recommendation model, while the outer optimization phase completes the denoising process guided by the accuracy of the inner recommendation model. Consequently, this approach primarily emphasizes the improvement of accuracy and thus releases the dependence on prior knowledge.

While the bi-level optimization approach to denoising has shown promise in addressing the limitations of existing methods, it has introduced new challenges as well. (1) It is storage-intensive because the size of the weights equals to the product of the user number and the item number in recommender systems if all samples are considered. As recommender systems expand, the number of users/items can become substantial, leading to a requirement for a large amount of memory to store the weights. (2) It is time-consuming to solve the bi-level optimization problem as conventional solutions require full training of the inner optimization to produce  suitable gradients for the outer optimization. To tackle these challenges, we provide an efficient \underline{b}i-level \underline{o}ptimization framework for \underline{d}enoising ($\textbf{BOD}$). For alleviating the storage demand, $\mathsf{BOD}$ employs the autoencoder network \cite{64kingma2013auto} based generator to generate the weights on the fly. This approach helps circumvent the high memory cost by storing the generator, which possesses fewer parameters in comparison to the weight variables. For reducing the computational time, $\mathsf{BOD}$ adopts a one-step update strategy to optimize the outer task instead of accumulating gradients from the inner optimization. Particularly, we employ the gradient matching technique \cite{19jin2021graph} to stabilize the training and ensure the validity of the gradient information in the one-step update. 

Our contributions are summarized as follows.
\begin{itemize}[leftmargin=12pt]
    \item We propose a bi-level optimization framework to tackle the noise in implicit feedback, which is prior knowledge-free and fully utilizes the varying sample weights throughout model training.
    \item We provide an efficient solution for the proposed bi-level optimization framework, which significantly reduces the storage demand and computational time.
    \item The experimental results on three benchmark datasets demonstrate that our proposed approach outperforms both state-of-the-art general and denoising recommendation models.
\end{itemize}

The rest of this paper is structured as follows. Section 2 provides the background on the relevant preliminaries. Section 3 presents the bi-level optimization framework for recommendation denoising. Section 4 provides the efficient solution to the proposed bi-level optimization problem. The experimental results and analysis are presented in Section 5. Section 6 reviews the related work for recommendation denoising. The paper is concluded in Section 7.

\section{Preliminary}
\subsection{Implicit Feedback Based Recommendation}
Given a user-item interaction data $\mathcal D=\{u,  i,  r_{u,i} | u \in \mathcal{U},i \in \mathcal{I}\}$, where $ \mathcal U\in\mathbb{R}^{|\mathcal U|}$ denotes the set of users, $ \mathcal I\in\mathbb{R}^{|\mathcal I|}$ denotes the set of items, and $r_{u,i}=\{0,1\}$ indicates whether user $u$ has interacted with item $i$. 
In general, recommendation methods based on implicit feedback are trained on interaction data $\mathcal D$, to learn user representations $\mathbf{Z}_{\mathcal U}\in\mathbb R^{|\mathcal U|\times d}$ ($d$ is the dimension of representations), item representations $\mathbf{Z}_{\mathcal I}\in\mathbb R^{|\mathcal I|\times d}$ and a model $f$ with parameters $\theta_{f}$ to make predictions.

The training of recommendation model is formulated as follows:

\begin{equation}
	\setlength{\abovedisplayskip}{-5pt}
	\setlength{\belowdisplayskip}{-1pt}
	\label{equation0}
	\begin{split}
		\quad \theta_{f}^{*}=\mathop{\arg\min}\limits_{\theta_{f}}  \mathcal{L}_{rec}(\mathcal{D}),
	\end{split}
\end{equation}
where $\mathcal{L}_{rec}$ is the recommendation loss, such as BPR loss \cite{28pan2013gbpr, 20rendle2014bayesian,21zhao2014leveraging} and AU loss \cite{17wang2022towards}, and $\theta_{f}^{*}$ is the optimal parameters of $f$. 
Here, we use the BPR loss as an instantiation of $L_{rec}$:
\begin{equation}
	\label{equation1}
	\begin{split}
		\mathcal{L}_{rec} = \underset{(u, i, j) \sim P_{\mathcal{D}}} {\mathbb{E}} log( \sigma(f({\mathbf{z}_{u}})^{T}f(\mathbf{z}_{i}))-f({\mathbf{z}_{u}})^{T}f(\mathbf{z}_{j})))),
	\end{split}
\end{equation}
where $P_{\mathcal{D}}(\cdot)$ refers to the distribution defined on the interaction data, the tuple $(u, i, j)$ denotes user $u$, a positive sample item $i$ with observed interactions, and a negative sample item $j$ without observed interactions with $u$.
This triple is obtained through the pairwise sampling of positive and negative samples of $u \in \mathcal{U}$ following $P_{\mathcal{D}}(\cdot)$, and $\sigma$ is the sigmoid function.

\subsection{ Recommendation Denoising}
Implicit feedback possesses a high degree of noise. The common way to denoise implicit feedback is to search a weight set $\mathcal W=\{u,  i,  w_{u,i} | u \in \mathcal{U},i \in \mathcal{I}\}$, where $0\leq w_{u,i} \leq 1$ indicates the weight for a sample (corresponds to the interaction between a user $u$ and an item $i$), which can reflect the probability of a sample being truly positive.
If the weights are known, the recommendation model is trained by considering the weights on the samples. 
Taking the weighted BRP loss as an instance:
\begin{equation}\label{equo:adt}
	\begin{split}
		\mathcal{L}_{rec} = \underset{(u, i, j) \sim P_{\mathcal{D}}} {\mathbb{E}} &log( \sigma(w_{u,i}f({\mathbf{z}_{u}})^{T}f(\mathbf{z}_{i}))
		\\&-w_{u,j}f({\mathbf{z}_{u}})^{T}f(\mathbf{z}_{j})))).
	\end{split}
\end{equation}
Existing denoising methods typically follow an iterative process of training a recommendation model with current weights and Equation~\ref{equo:adt}, followed by weight updates based on the recommendation model's loss. As discussed in Section 1, this process has two limitations: the neglect of previous weight information and the reliance on prior knowledge to identify potential noisy samples. To address these limitations, we introduce a novel denoising method based on bi-level optimization.

\section{Bi-Level Optimization for Denoising}
Under the bi-level optimization setting, the inner optimization is to derive the recommendation model $f$, and the outer optimization is to refine the sample weight variable $\mathcal{W}$. The bi-level optimization is formulated as follows:
\begin{equation}
\label{equation4}
\min_{\mathcal W}\mathcal{L}_{\mathcal W}(f_{\theta_{f}^{*}}), \quad s.t., \theta_{f}^{*}=\mathop{\arg\min}\limits_{\theta_{f}}\mathcal{L}_{rec}(\mathcal{W},\mathcal{D}),
\end{equation}
where $\mathcal{L}_{\mathcal{W}}$ is a defined loss for optimizing $\mathcal W$, and $\mathcal{L}_{rec}$ is a recommendation loss over interaction data $\mathcal{D}$ with weights $\mathcal{W}$. When optimizing $\mathcal W$ with $\mathcal{L}_{\mathcal{W}}$, we fix the recommendation model's current parameters $\theta_{f}$. Analogously, we fix the current weights when optimizing the recommendation model $f$ with $\mathcal{L}_{rec}$. In this framework, the weight set $\mathcal{W}$ is a learnable global variable. Therefore, the adjustments of $\mathcal{W}$ in previous iterations affect the computation of the current $\mathcal{W}$, which avoids the limitation that existing methods cannot share weight information throughout model training. In addition, the inner optimization guides the outer optimization for learning appropriate weight variable $\mathcal{W}$, eliminating the need for prior knowledge.

Despite the benefits of bi-level optimization, it also comes with new challenges. As stated in Section 1, in our setting it is infeasible to store $\mathcal{W}$ due to its sheer size, which is equal to $|\mathcal U|\times|\mathcal{I}|$ when all possible interactions are considered. Our empirical findings indicate that explicitly storing $\mathcal{W}$ for even moderate-sized datasets can cause out-of-memory issues. Besides, an intuitive approach to solving the bi-level optimization is to use the gradient information generated in the inner optimization to guide the learning of $\mathcal{W}$ in the outer optimization \cite{46zugner2018adversarial}, which requires the full training of the inner optimization to obtain suitable gradients for the outer optimization, leading to a significant time cost.

\section{Solving Bi-Level Optimization}
In this section, we provide an efficient solution to the bi-level optimization problem in Section 3.

\subsection{Generator-Based Parameter Generation}
\noindent

To reduce the storage demand of $\mathcal{W}$, we apply a generative network to generate each entry $w_{u,i}$ in $\mathcal{W}$. Equation \ref{equation2} shows the generator as follows: 
\begin{equation}
\label{equation2}
\begin{split}
w_{u,i} = g(f_{\theta_{f}^{*}}(\mathbf{z}_{u}),f_{\theta_{f}^{*}}(\mathbf{z}_{i})),
\end{split}
\end{equation}
where $g$ is the function of the generator whose parameters are $\theta_{g}$. The input of $g$ is the user embedding $\mathbf{z}_{u}$ and the item embedding $\mathbf{z}_{i}$ (transformed by the recommendation model $f_{\theta_{f}^{*}}$), and the output of $g$ is the corresponding $w_{u,i}$. 

The network architecture of $g$ is a simplified version of the AE (AutoEncoder) network \cite{45rifai2011contractive}. As per the conventional AE network, it comprises two essential components, namely the encoder layer and the decoder layer. The formulation of $g$ is outlined as follows:
\begin{align}
\label{equation6-7}
&\mathbf{z}=g_{e}(concat(f_{\theta_{f}^{*}}(\mathbf{z}_{u}),f_{\theta_{f}^{*}}(\mathbf{z}_{i}))), \\
&w_{u,i}=g_{d}(\mathbf{z}), 
\end{align}
where $g_{e}$ and $g_{d}$ refer to the fully connected layers of the encoder and decoder, respectively. In an effort to reduce the number of network parameters, we exclusively employ 1-lay fully connected layer as the encoder and decoder layers.

As demonstrated by Equation \ref{equation2}, the weight $w_{u,i}$ is not solely dependent on an update to the generator $g$, but is contingent upon the interplay between the generator $g$, the optimized recommendation model $f_{\theta_{f}^{*}}$, the user representation $\mathbf{z}_{u}$, and the item representation $\mathbf{z}_{i}$. This design is efficient because of two reasons: (1) \underline{Great scalability and flexibility}. The generator unifies all $w_{u,i}$ in $\mathcal{W}$. The input $\mathbf{z}_{u}$ and $\mathbf{z}_{i}$ encoded by $f_{\theta_{f}^{*}}$ can guarantee that $w_{u,i}$ becomes close if two user-item pairs are similar in the latent space, and the generator $g$ provides $w_{u,i}$ more variation.
(2) \underline{Low space cost}. Despite the additional training time of the generator, this design can avoid a significant space cost. More analysis of the model size and time complexity can be seen in Section \ref{section_modelanalysis}.

After using the generator, the update of $\mathcal{W}$ in outer optimization changes to the training of generator $g$, and the bi-level optimization changes as follows:
\begin{equation}
\label{equation5}
\min_{\theta_{g}}\mathcal{L}_{\mathcal W}(f_{\theta_{f}^{*}}), \quad s.t.,\ \  \theta_{f}^{*}=\mathop{\arg\min}\limits_{\theta_{f}} \mathcal{L}_{rec}
\end{equation}
\subsection{One-Step Gradient-Matching-Based Solving}
To reduce the high time cost, in this part, we propose a one-step gradient matching scheme. Specifically, we perform only a single update to the recommendation model within the inner optimization. To ensure that the gradient information from a single update can effectively contribute to the outer optimization, we introduce a gradient-matching mechanism that matches the gradient information of the two models during the inner optimization.

This idea builds upon a prior work~\cite{46zugner2018adversarial} that addresses the bi-level optimization using gradient information from multiple updates. During the inner optimization, the recommendation model is optimized using gradient descent with a learning rate $\eta$, shown as:
\begin{equation}
\label{equation_decent}
\begin{split}
\theta_{f_{t+1}} \leftarrow \theta_{f_t}-\eta\nabla_{\theta_{f_t}}\mathcal{L}_{rec}, 
\end{split}
\end{equation}
where $\nabla_{\theta_{f_t}}\mathcal{L}_{rec}$ represents the $t$-step gradient of recommendation model $f$ with regard to $\mathcal{L}_{rec}$, and multiple updates are allowed. In the outer optimization, it is required to unroll the whole training trajectory of the inner problem, which is computationally challenging and hinders scalability, leading us to develop an efficient optimization technique.

With the aim to reduce the high time cost associated with the optimization process, it is logical to think about updating the parameters in the inner optimization only once:
in Equation~\ref{equation_decent}, only one gradient $\nabla_{\theta_{f}}\mathcal{L}_{rec}$ is generated, and the parameter $\theta_{f}$ is updated immediately.
We thus remove the step notation $t$ from Equation~\ref{equation_decent} since we only perform the one-step calculation.
After obtaining the (one-step) gradient $\nabla_{\theta_{f}}\mathcal{L}_{rec}$ and update $\theta_{f}$, we proceed to update $\theta_{g}$ based on gradient $\nabla_{\theta_{f}}\mathcal{L}_{rec}$. This leads to an efficient optimization, as it avoids the need for multiple updates \textit{w.r.t} parameters in the inner optimization, as required by the outer optimization.

However, while the one-step gradient leading training accelerates the computational process, it may negatively impact the training of the generator $g$, as it reduces the amount of gradient information available. Multiple gradients contain more valuable information, such as the dynamic trend of the gradient, which is necessary for guiding the training of $g$ in the correct direction. To tackle this issue, we propose to use additional recommendation losses to train recommendation model simultaneously, which provide more gradient information. It is important to note that blindly incorporating all gradient information into the training of $g$ would not result in a positive change. Instead, an elegant combination of the gradient information is necessary to effectively train $g$.

Motivated by \cite{04nguyen2021dataset, 18zhao2021dataset, 19jin2021graph, 53jin2022condensing}, we adopt the gradient matching scheme to guide the optimization of the generator $g$. Specifically, gradient matching involves defining two types of losses and optimizing the model by minimizing the difference between them, i.e.,
the models trained on these two losses converge to similar parameters. With the gradient matching scheme, the ultimate form of the bi-level optimization is as follows:
\begin{equation}
\label{equation8}
\begin{split}
&\min_{\theta_{g}}D(\nabla_{\theta_{ft}}\mathcal{L}_{rec1},\nabla_{\theta_{ft}}\mathcal{L}_{rec2}), \\ & s.t.,  \theta_{f}^{*}=\mathop{\arg\min}\limits_{\theta_{f}} (\mathcal{L}_{rec1}(\mathcal{W},\mathcal{D})+\alpha\mathcal{L}_{rec2}(\mathcal{W},\mathcal{D})),
\end{split}
\end{equation}
where weight $\alpha$ control the balance of two recommendation loss, and $\mathcal{L}_{rec1}$ and $\mathcal{L}_{rec2}$ can be any two common recommendation loss functions. $\nabla_{\theta_{ft}}\mathcal{L}_{rec1}$ and $\nabla_{\theta_{ft}}\mathcal{L}_{rec2}$ are gradients of $\mathcal{L}_{rec1}$ and $\mathcal{L}_{rec2}$ respectively. In the gradient matching, our purpose is to optimize $g$ such that two different one-step gradients are pulled closer, and the $D(\cdot)$ is the distance function \cite{19jin2021graph}, which is shown as follows:
\vspace{-0.5em}
\begin{equation}
\label{equation9}
\begin{split}
D(G^{1}, G^{2})= \sum_{c=1}^{d_{2}}(1-\frac{G^{1}_{c} \cdot G^{2}_{c}}{\Vert G^{1}_{c} \Vert\Vert G^{2}_{c} \Vert}),
\end{split}
\end{equation}
where $G^{1}\in\mathbb{R}^{d_{1} \times d_{2}}$ and $G^{2}\in\mathbb{R}^{d_{1} \times d_{2}}$ are gradients at a specific layer, $d_{1}$, $d_{2}$ are the number of rows and columns of the gradient matrices, and $G^{1}_{c} \in \mathbb{R}^{d_1}$ and $G^{2}_{c} \in \mathbb{R}^{d_1}$ refer to the $c$-th column vectors of the gradient matrices. 

The advantages of using the gradient matching scheme include: (1) \underline{Diverse perspectives.} Gradient matching entails utilizing two distinct gradients, enabling the model to update and explore the data from various angles, thereby enhancing its performance. (2) \underline{Improved stability.} By aligning or matching the different gradients, the model's training process can potentially remain stable. This alignment ensures reasonable exploitation of the data. In clean samples, the gradients corresponding to the two losses exhibit minimal differences and remain stable. Conversely, noisy samples may result in substantial differences between the gradients. By aligning the two gradients, the scheme effectively emphasizes the differences between clean and noisy samples, further highlighting the disparity between them. (3) \underline{Efficiency in computational cost.} The use of one-step gradient information during the optimization of the generator $g$ reduces the computational overhead as it is generated during the training of the recommendation model;

Note that, since $\nabla_{\theta_{ft}}\mathcal{L}_{rec1}$ and $\nabla_{\theta_{ft}}\mathcal{L}_{rec2}$ are unstable in the initial stage of recommendation model training, the one-step gradient matching might fail if directly using to train generator $g$ because of the large difference between $\nabla_{\theta_{ft}}\mathcal{L}_{rec1}$ and $\nabla_{\theta_{ft}}\mathcal{L}_{rec2}$. To get a relatively suitable $\nabla_{\theta_{ft}}\mathcal{L}_{rec1}$ and $\nabla_{\theta_{ft}}\mathcal{L}_{rec2}$, we will pre-train the recommendation model $f$ to a stable state, i.e., the warm-up stage, and then use one-step gradient matching to train $g$.

\subsection{The Framework BOD}

\begin{figure}[h]
  \centering
  \includegraphics[width=0.85\linewidth]{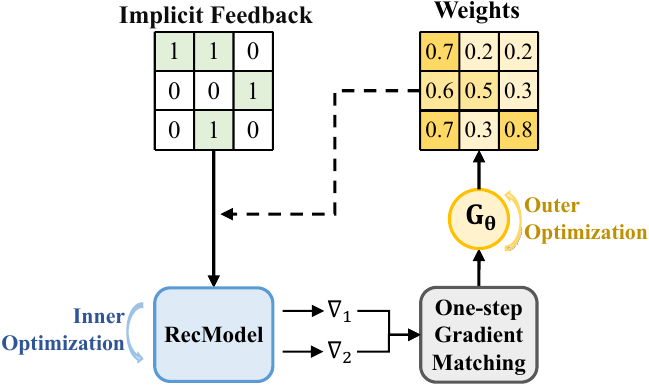}
  \caption{The framework $\mathsf{BOD}$.}
  \vspace{-1em}
\label{method}
\end{figure}

By incorporating the concept of generator-based parameter generation and one-step gradient matching scheme, we effectively minimize the computational demands of bi-level optimization for denoising. The resulting framework, referred to as $\mathsf{BOD}$ (as depicted in Fig.~\ref{method}), serves as a general solution for recommendation denoising. Algorithm ~\ref{algorithm1} shows the learning process of $\mathsf{BOD}$. It is described as follows: (1) \underline{Towards inner optimization solving (Line~3-6)}: the inner optimization is to train the recommendation model over implicit feedback data with weights. Using two losses, we can generate corresponding one-step gradients $\nabla_{\theta_{ft}^{rec1}}$ and $\nabla_{\theta_{ft}^{rec2}}$. (2) \underline{Towards outer optimization solving (Line~7-8)}: the outer optimization uses gradient matching to guide the generator training, which can generate weights for the next step of bi-level optimization.

\begin{algorithm}[h]
		\caption{The proposed $\mathsf{BOD}$ framework}
		\label{algorithm1}
        \KwIn{interactions data $\mathcal{D}$, recommendation model $f$, generator $g$}
        \KwOut{recommendation model with optimal para. $f_{\theta_{f}^{*}}$}
  
        warm-up the recommendation model $f$ and randomly initialize parameters of $g$\;
        \While{not converged}
        {
        \tcp{inner optimization: fix $g$'s parameters $\theta_{g}$}
        sample a batch tuple $(u,i,j)$, and generate corresponding $w_{u,i}$ and $w_{u,j}$ by $g$\;
        calculate $\mathcal{L}_{rec1}$ and $\mathcal{L}_{rec2}$\;
        one update $f_{\theta_{f}}$ to $f_{\theta_{f}^{*}}$ based on $\mathcal{L}_{rec1}$+$\alpha\mathcal{L}_{rec2}$\;
        record $\nabla_{\theta_{ft}^{rec1}}$ and $\nabla_{\theta_{ft}^{rec2}}$\;
        \tcp{outer optimization:fix $f$'s optimal parameters $\theta_{f}^{*}$}
        calculate matching distance loss $D(\nabla_{\theta_{ft}^{rec1}}, \nabla_{\theta_{ft}^{rec2}}$)\;
        optimize $\theta_{g}$ based on $D(\nabla_{\theta_{ft}^{rec1}}, \nabla_{\theta_{ft}^{rec2}})$\;
        }
		
\end{algorithm} 

The framework $\mathsf{BOD}$ is characterized by its generality, as demonstrated in Equation~\ref{equation8}, where the specific recommendation model is not specified. To apply the framework to a specific recommendation model $f$, the two recommendation losses for gradient matching must be specified, as described in Algorithm~\ref{algorithm1}. The choice of the recommendation losses is flexible and not subject to any constraints. As a demonstration, we choose the BRP loss \cite{28pan2013gbpr, 20rendle2014bayesian,21zhao2014leveraging}, which is a classical method used in recommendation denoising (referred to in Section 2), and the AU loss \cite{17wang2022towards} (presented in Appendix \ref{wau}), which is a recent development. Through this example, the denoising method for the specific recommendation model $f$ can be generated based on the $\mathsf{BOD}$ framework and the two loss functions specified. Our contribution to the field is not the specific choice of recommendation losses, but rather the design of a general denoising framework $\mathsf{BOD}$ that can incorporate any two recommendation losses for recommendation denoising.

\subsection{Complexity Analysis}
\label{section_modelanalysis}
We summarize the complexity of the base model, existing denoising methods, $\mathsf{BOD}$ without the generator, and $\mathsf{BOD}$ in Table \ref{model analysis}.

\stitle{Model Size.} The parameters of $\mathsf{BOD}$ come from two parts: (1) the parameters of recommendation models, which we assume is $M$; and (2) the parameters of the generator. For the generator, we need additional parameters for the encoder layer $EN\in\mathbb{R}^{2d \times d_{g}}$, and the decoder layer $DE\in\mathbb{R}^{d_{g} \times 1}$, where $d$ is the embedding size of user and item in recommendation model, and $d_{g}$ is the hidden layer size of the generator. Overall, the space cost of $\mathsf{BOD}$ is $M+EN+DE$, which is negligible compared with full weights matrix $\mathcal{W}\in\mathbb{R}^{|\mathcal U| \times |\mathcal I|}$. This shows that the space cost can be greatly reduced by using the generator.

\stitle{Time Complexity.} The complexity of $\mathsf{BOD}$ consists of two parts. (1) Inner optimization: the update of recommendation model parameters, which we assume is $\mathcal O(M)$; and (2) Outer optimization: the update of the generator, which generates weights. The time complexity of the generator includes encoder layer $\mathcal O(d (d_{g}))$ and the decoder layer $\mathcal O(d_{g})$, respectively. It is obvious that our scheme can save a lot of time costs because $\mathcal O(M|\mathcal U||\mathcal I|)$ appears in other denoising methods are much larger.

\begin{table}[h]
  \caption{Model complexity comparison.}\vspace{-1em}
  \small
  \label{model analysis}
\begin{tabular}{ccc}
\hline
\textbf{Model}            & \textbf{Model Size} & \textbf{Time Complexity} \\ \hline
Base Model                   & $M$          & $\mathcal O(M)$               \\ 
\makecell{Denoising methods} & $M+|\mathcal U||\mathcal I|$          & $\mathcal O(M |\mathcal U||\mathcal I|)$               \\ 
$\mathsf{BOD}\ w/o\ gen$             & $M+|\mathcal U||\mathcal I|$          & $\mathcal O(M +|\mathcal U||\mathcal I|)$                        \\ 
$\mathsf{BOD}$                & $M+EN+DE$         & $\mathcal O(M + d (d_{g}) + (d_{g}))$              \\ \hline
\end{tabular}
\end{table}

\section{Experiments}
This section tests the effectiveness of our proposed $\mathsf{BOD}$.
Specifically, we aim to answer the following questions.(RQ1): How does the performance and robustness of $\mathsf{BOD}$ against cutting-edge denoising methods? (RQ2): What is the running time of $\mathsf{BOD}$? (RQ3): How does the inner optimization affect $\mathsf{BOD}$? (RQ4): How does the generator affect $\mathsf{BOD}$? (RQ5): What is the stability of $\mathsf{BOD}$? (RQ6): How about the generalizability of $\mathsf{BOD}$?

\begin{table}[h]
\caption{Statistics of datasets.}\vspace{-1em}
\small
\label{datasets}
\begin{tabular}{@{}ccccc@{}}
\toprule
\textbf{Dataset}                                            & \textbf{\#Users} & \textbf{\#Items} & \textbf{\#Interactions} & \textbf{Density} \\ \midrule
Beauty                                                      & 22,363           & 12,099           & 198,503                 & 0.073\%          \\
\begin{tabular}[c]{@{}c@{}}iFashion\end{tabular} & 300,000          & 81,614           & 1,607,813               & 0.006\%          \\
Yelp2018                                                    & 31,668           & 38,048           & 1,561,406               & 0.130\%          \\ \bottomrule
\end{tabular}
\end{table}

\vspace{-1em}
\subsection{Performance and Robustness (RQ1)}

\begin{table*}[h]
\caption{Performance comparison of different denoising methods on the robust recommendation. The highest scores are in bold, and the second best are with underlines. R and N refer to Recall and NDCG, respectively.}\vspace{-1em}
\centering
\resizebox{0.9\textwidth}{!}{
\begin{tabular}{@{}cc|llll|llll|llll@{}}
\toprule
\multicolumn{2}{c|}{Dataset}                            & \multicolumn{4}{c|}{Beauty}                                                                                                                    & \multicolumn{4}{c|}{iFashion}                                                                                                          & \multicolumn{4}{c}{Yelp2018}                                                                                                                  \\ \midrule
\multicolumn{1}{c|}{Base Model}                & Method & \multicolumn{1}{c|}{R@10}            & \multicolumn{1}{c|}{R@20}            & \multicolumn{1}{c|}{N@10}            & \multicolumn{1}{c|}{N@20} & \multicolumn{1}{c|}{R@10}            & \multicolumn{1}{c|}{R@20}            & \multicolumn{1}{c|}{N@10}            & \multicolumn{1}{c|}{N@20} & \multicolumn{1}{c|}{R@10}            & \multicolumn{1}{c|}{R@20}            & \multicolumn{1}{c|}{N@10}            & \multicolumn{1}{c}{N@20} \\ \midrule
\multicolumn{1}{c|}{\multirow{7}{*}{GMF}}      & Normal & \multicolumn{1}{l|}{0.0639}          & \multicolumn{1}{l|}{0.0922}          & \multicolumn{1}{l|}{0.0429}          & 0.0518                    & \multicolumn{1}{l|}{0.0286}          & \multicolumn{1}{l|}{0.0462}          & \multicolumn{1}{l|}{0.0157}          & 0.0205                    & \multicolumn{1}{l|}{0.0290}          & \multicolumn{1}{l|}{0.0500}          & \multicolumn{1}{l|}{0.0331}          & 0.0411                   \\
\multicolumn{1}{c|}{}                          & WBPR   & \multicolumn{1}{l|}{0.0640}          & \multicolumn{1}{l|}{0.0919}          & \multicolumn{1}{l|}{0.0430}          & 0.0516                    & \multicolumn{1}{l|}{0.0284}          & \multicolumn{1}{l|}{0.0461}          & \multicolumn{1}{l|}{0.0158}          & 0.0205                    & \multicolumn{1}{l|}{0.0289}          & \multicolumn{1}{l|}{0.0498}          & \multicolumn{1}{l|}{0.0330}          & 0.0409                   \\
\multicolumn{1}{c|}{}                          & WRMF & \multicolumn{1}{l|}{0.0714}          & \multicolumn{1}{l|}{0.1039}          & \multicolumn{1}{l|}{0.0480}          & 0.0582                    & \multicolumn{1}{l|}{0.0397}          & \multicolumn{1}{l|}{0.0589}          & \multicolumn{1}{l|}{0.0211}          & 0.0265                    & \multicolumn{1}{l|}{0.0304}          & \multicolumn{1}{l|}{0.0539}          & \multicolumn{1}{l|}{0.0338}          & 0.0422                   \\
\multicolumn{1}{c|}{}                          & T-CE   & \multicolumn{1}{l|}{0.0785}          & \multicolumn{1}{l|}{0.1085}          & \multicolumn{1}{l|}{0.0532}          & 0.0654                    & \multicolumn{1}{l|}{0.0385}          & \multicolumn{1}{l|}{0.0583}          & \multicolumn{1}{l|}{0.0193}          & 0.0256                    & \multicolumn{1}{l|}{0.0302}          & \multicolumn{1}{l|}{0.0538}          & \multicolumn{1}{l|}{0.0335}          & 0.0421                   \\
\multicolumn{1}{c|}{}                          & DeCA   & \multicolumn{1}{l|}{0.0767}          & \multicolumn{1}{l|}{0.1081}          & \multicolumn{1}{l|}{0.0535}          & 0.0669                    & \multicolumn{1}{l|}{0.0426}          & \multicolumn{1}{l|}{0.0626}          & \multicolumn{1}{l|}{{\ul 0.0234}}    & {\ul 0.0297}              & \multicolumn{1}{l|}{0.0298}          & \multicolumn{1}{l|}{0.0532}          & \multicolumn{1}{l|}{0.0336}          & 0.0415                   \\
\multicolumn{1}{c|}{}                          & SGDL   & \multicolumn{1}{l|}{{\ul 0.0804}}    & \multicolumn{1}{l|}{{\ul 0.1092}}    & \multicolumn{1}{l|}{{\ul 0.0546}}    & {\ul 0.0675}              & \multicolumn{1}{l|}{{\ul 0.0435}}    & \multicolumn{1}{l|}{{\ul 0.0646}}    & \multicolumn{1}{l|}{0.0226}          & 0.0289                    & \multicolumn{1}{l|}{{\ul 0.0305}}    & \multicolumn{1}{l|}{{\ul 0.0541}}    & \multicolumn{1}{l|}{{\ul 0.0351}}    & {\ul 0.0436}             \\ \cmidrule(l){2-14} 
\multicolumn{1}{c|}{}                          & $\mathsf{BOD}$    & \multicolumn{1}{l|}{\textbf{0.0843}} & \multicolumn{1}{l|}{\textbf{0.1193}} & \multicolumn{1}{l|}{\textbf{0.0579}} & \textbf{0.0688}           & \multicolumn{1}{l|}{\textbf{0.0596}} & \multicolumn{1}{l|}{\textbf{0.0860}} & \multicolumn{1}{l|}{\textbf{0.0355}} & \textbf{0.0426}           & \multicolumn{1}{l|}{\textbf{0.0357}} & \multicolumn{1}{l|}{\textbf{0.0621}} & \multicolumn{1}{l|}{\textbf{0.0408}} & \textbf{0.0513}          \\ \midrule
\multicolumn{1}{c|}{\multirow{7}{*}{NCF}}      & Normal & \multicolumn{1}{l|}{0.0738}          & \multicolumn{1}{l|}{0.1075}          & \multicolumn{1}{l|}{0.0488}          & 0.0593                    & \multicolumn{1}{l|}{0.0404}          & \multicolumn{1}{l|}{0.0632}          & \multicolumn{1}{l|}{0.0227}          & 0.0288                    & \multicolumn{1}{l|}{0.0281}          & \multicolumn{1}{l|}{0.0493}          & \multicolumn{1}{l|}{0.0324}          & 0.0406                   \\
\multicolumn{1}{c|}{}                          & WBPR   & \multicolumn{1}{l|}{0.0741}          & \multicolumn{1}{l|}{0.1082}          & \multicolumn{1}{l|}{0.0494}          & 0.0600                    & \multicolumn{1}{l|}{0.0410}          & \multicolumn{1}{l|}{0.0638}          & \multicolumn{1}{l|}{0.0231}          & 0.0292                    & \multicolumn{1}{l|}{0.0288}          & \multicolumn{1}{l|}{0.0499}          & \multicolumn{1}{l|}{0.0331}          & 0.0410                   \\
\multicolumn{1}{c|}{}                          & WRMF   & \multicolumn{1}{l|}{0.0745}          & \multicolumn{1}{l|}{0.1097}          & \multicolumn{1}{l|}{0.0498}          & 0.0611                    & \multicolumn{1}{l|}{0.0455}          & \multicolumn{1}{l|}{0.0667}          & \multicolumn{1}{l|}{0.0247}          & 0.0306                    & \multicolumn{1}{l|}{0.0292}          & \multicolumn{1}{l|}{0.0503}          & \multicolumn{1}{l|}{0.0335}          & 0.0413                   \\
\multicolumn{1}{c|}{}                          & T-CE   & \multicolumn{1}{l|}{0.0798}          & \multicolumn{1}{l|}{0.1123}          & \multicolumn{1}{l|}{0.0506}          & 0.0632                    & \multicolumn{1}{l|}{0.0546}          & \multicolumn{1}{l|}{0.0684}          & \multicolumn{1}{l|}{0.0268}          & 0.0368                    & \multicolumn{1}{l|}{0.0301}          & \multicolumn{1}{l|}{0.0521}          & \multicolumn{1}{l|}{0.0335}          & 0.0415                   \\
\multicolumn{1}{c|}{}                          & DeCA   & \multicolumn{1}{l|}{{\ul 0.0886}}    & \multicolumn{1}{l|}{{\ul 0.1265}}    & \multicolumn{1}{l|}{0.0588}          & 0.0646                    & \multicolumn{1}{l|}{0.0547}          & \multicolumn{1}{l|}{0.0756}          & \multicolumn{1}{l|}{{\ul 0.0315}}    & {\ul 0.0421}              & \multicolumn{1}{l|}{0.0297}          & \multicolumn{1}{l|}{0.0535}          & \multicolumn{1}{l|}{0.0351}          & 0.0459                   \\
\multicolumn{1}{c|}{}                          & SGDL   & \multicolumn{1}{l|}{0.0864}          & \multicolumn{1}{l|}{0.1235}          & \multicolumn{1}{l|}{{\ul 0.0610}}    & {\ul 0.0698}              & \multicolumn{1}{l|}{{\ul 0.0588}}    & \multicolumn{1}{l|}{{\ul 0.0846}}    & \multicolumn{1}{l|}{0.0311}          & 0.0329                    & \multicolumn{1}{l|}{{\ul 0.0325}}    & \multicolumn{1}{l|}{{\ul 0.0598}}    & \multicolumn{1}{l|}{{\ul 0.0388}}    & {\ul 0.0465}             \\ \cmidrule(l){2-14} 
\multicolumn{1}{c|}{}                          & $\mathsf{BOD}$    & \multicolumn{1}{l|}{\textbf{0.0959}} & \multicolumn{1}{l|}{\textbf{0.1352}} & \multicolumn{1}{l|}{\textbf{0.0652}} & \textbf{0.0774}           & \multicolumn{1}{l|}{\textbf{0.0624}} & \multicolumn{1}{l|}{\textbf{0.0909}} & \multicolumn{1}{l|}{\textbf{0.0366}} & \textbf{0.0442}           & \multicolumn{1}{l|}{\textbf{0.0370}} & \multicolumn{1}{l|}{\textbf{0.0633}} & \multicolumn{1}{l|}{\textbf{0.0434}} & \textbf{0.0531}          \\ \midrule
\multicolumn{1}{c|}{\multirow{5}{*}{NGCF}}     & Normal & \multicolumn{1}{l|}{0.0726}          & \multicolumn{1}{l|}{0.1080}          & \multicolumn{1}{l|}{0.0465}          & 0.0575                    & \multicolumn{1}{l|}{0.0351}          & \multicolumn{1}{l|}{0.0579}          & \multicolumn{1}{l|}{0.0188}          & 0.0250                    & \multicolumn{1}{l|}{0.0231}          & \multicolumn{1}{l|}{0.0418}          & \multicolumn{1}{l|}{0.0263}          & 0.0336                   \\
\multicolumn{1}{c|}{}                          & WBPR   & \multicolumn{1}{l|}{0.0731}          & \multicolumn{1}{l|}{0.1092}          & \multicolumn{1}{l|}{0.0476}          & 0.0589                    & \multicolumn{1}{l|}{0.0364}          & \multicolumn{1}{l|}{0.0598}          & \multicolumn{1}{l|}{0.0197}          & 0.0266                    & \multicolumn{1}{l|}{0.0240}          & \multicolumn{1}{l|}{0.0422}          & \multicolumn{1}{l|}{0.0271}          & 0.0343                   \\
\multicolumn{1}{c|}{}                          & DeCA   & \multicolumn{1}{l|}{0.0834}          & \multicolumn{1}{l|}{0.1275}          & \multicolumn{1}{l|}{0.0587}          & 0.0665                    & \multicolumn{1}{l|}{{\ul 0.0587}}    & \multicolumn{1}{l|}{{\ul 0.0855}}    & \multicolumn{1}{l|}{{\ul 0.0247}}    & {\ul 0.0356}              & \multicolumn{1}{l|}{0.0287}          & \multicolumn{1}{l|}{0.0486}          & \multicolumn{1}{l|}{0.0299}          & 0.0435                   \\
\multicolumn{1}{c|}{}                          & SGDL   & \multicolumn{1}{l|}{{\ul 0.0935}}    & \multicolumn{1}{l|}{{\ul 0.1389}}    & \multicolumn{1}{l|}{{\ul 0.0635}}    & {\ul 0.0758}              & \multicolumn{1}{l|}{0.0566}          & \multicolumn{1}{l|}{0.0848}          & \multicolumn{1}{l|}{0.0225}          & 0.0345                    & \multicolumn{1}{l|}{{\ul 0.0297}}    & \multicolumn{1}{l|}{{\ul 0.0503}}    & \multicolumn{1}{l|}{{\ul 0.0348}}    & {\ul 0.0445}             \\ \cmidrule(l){2-14} 
\multicolumn{1}{c|}{}                          & $\mathsf{BOD}$    & \multicolumn{1}{l|}{\textbf{0.1015}} & \multicolumn{1}{l|}{\textbf{0.1415}} & \multicolumn{1}{l|}{\textbf{0.0702}} & \textbf{0.0827}           & \multicolumn{1}{l|}{\textbf{0.0716}} & \multicolumn{1}{l|}{\textbf{0.1053}} & \multicolumn{1}{l|}{\textbf{0.0390}} & \textbf{0.0480}           & \multicolumn{1}{l|}{\textbf{0.0331}} & \multicolumn{1}{l|}{\textbf{0.0565}} & \multicolumn{1}{l|}{\textbf{0.0377}} & \textbf{0.0466}          \\ \midrule
\multicolumn{1}{c|}{\multirow{7}{*}{LightGCN}} & Normal & \multicolumn{1}{l|}{0.0855}          & \multicolumn{1}{l|}{0.1221}          & \multicolumn{1}{l|}{0.0561}          & 0.0675                    & \multicolumn{1}{l|}{0.0429}          & \multicolumn{1}{l|}{0.0661}          & \multicolumn{1}{l|}{0.0247}          & 0.0310                    & \multicolumn{1}{l|}{0.0308}          & \multicolumn{1}{l|}{0.0532}          & \multicolumn{1}{l|}{0.0361}          & 0.0445                   \\
\multicolumn{1}{c|}{}                          & WBPR   & \multicolumn{1}{l|}{0.0864}          & \multicolumn{1}{l|}{0.1261}          & \multicolumn{1}{l|}{0.0588}          & 0.0685                    & \multicolumn{1}{l|}{0.0431}          & \multicolumn{1}{l|}{0.0662}          & \multicolumn{1}{l|}{0.0253}          & 0.0316                    & \multicolumn{1}{l|}{0.0317}          & \multicolumn{1}{l|}{0.0529}          & \multicolumn{1}{l|}{0.0365}          & 0.0448                   \\
\multicolumn{1}{c|}{}                          & DeCA   & \multicolumn{1}{l|}{0.0967}          & \multicolumn{1}{l|}{0.1345}          & \multicolumn{1}{l|}{0.0665}          & 0.0754                    & \multicolumn{1}{l|}{0.0540}          & \multicolumn{1}{l|}{0.0865}          & \multicolumn{1}{l|}{0.0354}          & 0.0435                    & \multicolumn{1}{l|}{0.0337}          & \multicolumn{1}{l|}{0.0511}          & \multicolumn{1}{l|}{0.0432}          & 0.0524                   \\
\multicolumn{1}{c|}{}                          & SGDL   & \multicolumn{1}{l|}{0.0946}          & \multicolumn{1}{l|}{0.1365}          & \multicolumn{1}{l|}{0.0677}          & 0.0769                    & \multicolumn{1}{l|}{0.0591}          & \multicolumn{1}{l|}{0.0908}          & \multicolumn{1}{l|}{0.0342}          & 0.0415                    & \multicolumn{1}{l|}{0.0339}          & \multicolumn{1}{l|}{0.0541}          & \multicolumn{1}{l|}{0.0441}          & 0.0575                   \\ \cmidrule(l){2-14} 
\multicolumn{1}{c|}{}                          & SGL    & \multicolumn{1}{l|}{{\ul 0.1005}}    & \multicolumn{1}{l|}{{\ul 0.1422}}    & \multicolumn{1}{l|}{{\ul 0.0692}}    & {\ul 0.0821}              & \multicolumn{1}{l|}{0.0665}          & \multicolumn{1}{l|}{0.0973}          & \multicolumn{1}{l|}{0.0392}          & 0.0475                    & \multicolumn{1}{l|}{0.0374}          & \multicolumn{1}{l|}{0.0639}          & \multicolumn{1}{l|}{0.0436}          & 0.0534                   \\
\multicolumn{1}{c|}{}                          & SimGCL & \multicolumn{1}{l|}{0.0960}          & \multicolumn{1}{l|}{0.1316}          & \multicolumn{1}{l|}{0.0671}          & 0.0781                    & \multicolumn{1}{l|}{{\ul 0.0738}}    & \multicolumn{1}{l|}{{\ul 0.1070}}    & \multicolumn{1}{l|}{{\ul 0.0434}}    & {\ul 0.0523}              & \multicolumn{1}{l|}{{\ul 0.0414}}    & \multicolumn{1}{l|}{\textbf{0.0711}} & \multicolumn{1}{l|}{{\ul 0.0485}}    & \textbf{0.0593}          \\ \cmidrule(l){2-14} 
\multicolumn{1}{c|}{}                          & $\mathsf{BOD}$    & \multicolumn{1}{l|}{\textbf{0.1095}} & \multicolumn{1}{l|}{\textbf{0.1548}} & \multicolumn{1}{l|}{\textbf{0.0755}} & \textbf{0.0895}           & \multicolumn{1}{l|}{\textbf{0.0811}} & \multicolumn{1}{l|}{\textbf{0.1180}} & \multicolumn{1}{l|}{\textbf{0.0477}} & \textbf{0.0576}           & \multicolumn{1}{l|}{\textbf{0.0416}} & \multicolumn{1}{l|}{{\ul 0.0700}}    & \multicolumn{1}{l|}{\textbf{0.0490}} & {\ul 0.0588}             \\ \bottomrule
\end{tabular}}
\label{performance comparison}
\end{table*}

\subsection{Experimental Settings}
\textbf{Datasets.} Three commonly used public benchmark datasets: Beauty \cite{17wang2022towards}, iFashion \cite{13wu2021self}, and Yelp2018 \cite{26he2020lightgcn}, are used in our experiments. The dataset statistics are shown in Table ~\ref{datasets}.

\stitle{Evaluation Metrics.} We split the datasets into three parts (training set, validation set, and test set) with a ratio of 7:1:2. Two common evaluation metrics are used, Recall$@K$ and NDCG$@K$. We set $K$=10 and $K$=20. Each metric is conducted 10 times, and then we report the average results. 

\stitle{Baselines.} The main objective of this paper is to denoise the feedback to improve the performance of recommender systems.
For this purpose, four commonly used (implicit feedback-based) recommendation models are chosen as \textbf{base models} for recommendation denoising: 
\begin{itemize}[leftmargin=12pt]
    \item GMF \cite{39he2017neural} is a generalized version of matrix factorization based recommendation model.
    \item NCF \cite{50he2017neural} generalizes collaborative filtering with a Multi-Layer Perceptron.
     \item NGCF \cite{51wang2019neural} applies graph convolution network (GCN) to encode user-item bipartite graph.
    \item LightGCN \cite{26he2020lightgcn} is a state-of-the-art graph model, which discards the nonlinear feature transformations to simplify the design of GCN for the recommendation.
\end{itemize}

\vspace{0.5em}
To compare the denoising effect, we first choose four existing denoising methods for the above recommendation models:
\begin{itemize}[leftmargin=12pt]
    \item WBPR \cite{08gantner2012personalized} considers popular but uninteracted items as true negative samples. We take the number of interactions of items as the popularity.
    \item WRMF \cite{60hu2008collaborative} uses weighted matrix factorization whose weights are fixed to denoise recommendation.
    \item T-CE \cite{11wang2021denoising} is the denoising method, which uses the binary cross-entropy (BCE) loss~\cite{22zhang2021double,23lian2020geography} to assign weights to large loss samples with a dynamic threshold.
    Because the BCE loss only applies to some recommendation models, we can only get (and thus report) the results of T-CE in GMF and NCF, but not in the other two base models.
    
    \item DeCA \cite{10wang2022learning} combines predictions of two different models to consider the disagreement of noisy samples.
    \item SGDL \cite{12DBLP:conf/sigir/Gao0HCZFZ22} is the state-of-the-art denoising model, which collects clean interactions in the initial training stages, and uses them to distinguish noisy samples based on the similarity of collected clean samples.
\end{itemize}

\vspace{0.5em}
To further confirm the effectiveness of our model, we also compare $\mathsf{BOD}$ with the state-of-the-art robust recommendation models SGL and SimGCL:
\begin{itemize}[leftmargin=12pt]
    \item SGL \cite{13wu2021self} applies edge dropout to modify discrete graph structure randomly and trains different graphs based on contrastive learning to enhance the representation of nodes.
    \item SimGCL \cite{15yu2022graph} simplifies the contrastive loss and optimizes representations' uniformity.
\end{itemize}

 Since SGL and SimGCL use LightGCN as the base model in their work, so we only compare them when LightGCN is used and ignore the results on the other base models. 

\stitle{Our Method $\mathsf{BOD}$.} If not stressed, the two loss functions of $\mathsf{BOD}$ are BPR and AU losses, respectively.

\stitle{Parameter Settings.} We implement $\mathsf{BOD}$ in Pytorch. If not stressed, we use recommended parameter settings for all models, in which batch size, learning rate, embedding size, and the number of LightGCN layers are 128, 0.001, 64, and 2, respectively. For SGL, the edge dropout rate is set to 0.1. We optimize them with Adam \cite{41kingma2014adam} optimizer and use the Xavier initializer \cite{42glorot2010understanding} to initialize the model parameters. 

\stitle{Performance Comparison.} We first compare $\mathsf{BOD}$ with existing denoising methods and robust recommendation methods on three different datasets. Table \ref{performance comparison} shows the results, where figures in bold represent the best performing indices, and the runner-ups are presented with underlines. According to Table \ref{performance comparison}, we can draw the following observations and conclusions:
\begin{itemize}[leftmargin=12pt]
    \item The proposed $\mathsf{BOD}$ can effectively improve the performance of four base models and show the best/second performance in all cases. We attribute these improvements to the storage and utilization of weights. In the bi-level process, $\mathsf{BOD}$ dynamically updates and stores the weights for all samples, including observed and unobserved samples. While other baselines (e.g., DeCA and SGDL) are insufficient to provide dynamically updated weights.
    \item All denoising approaches have better results than normal training, which indicates the validity of denoising in recommendations. The results are consistent with prior studies \cite{10wang2022learning,12DBLP:conf/sigir/Gao0HCZFZ22}.
    \item The improvement on Yelp2018 dataset is less significant than that on other datasets. One possible reason is that Yelp2018 dataset has a higher density than the others. Thus there are enough pure informative interactions for identifying user behavior, compensating for the effect of noisy negative samples.
\end{itemize}

\begin{figure}[t]
\setlength{\belowcaptionskip}{-15pt}
  \centering
  \includegraphics[width=1\linewidth]{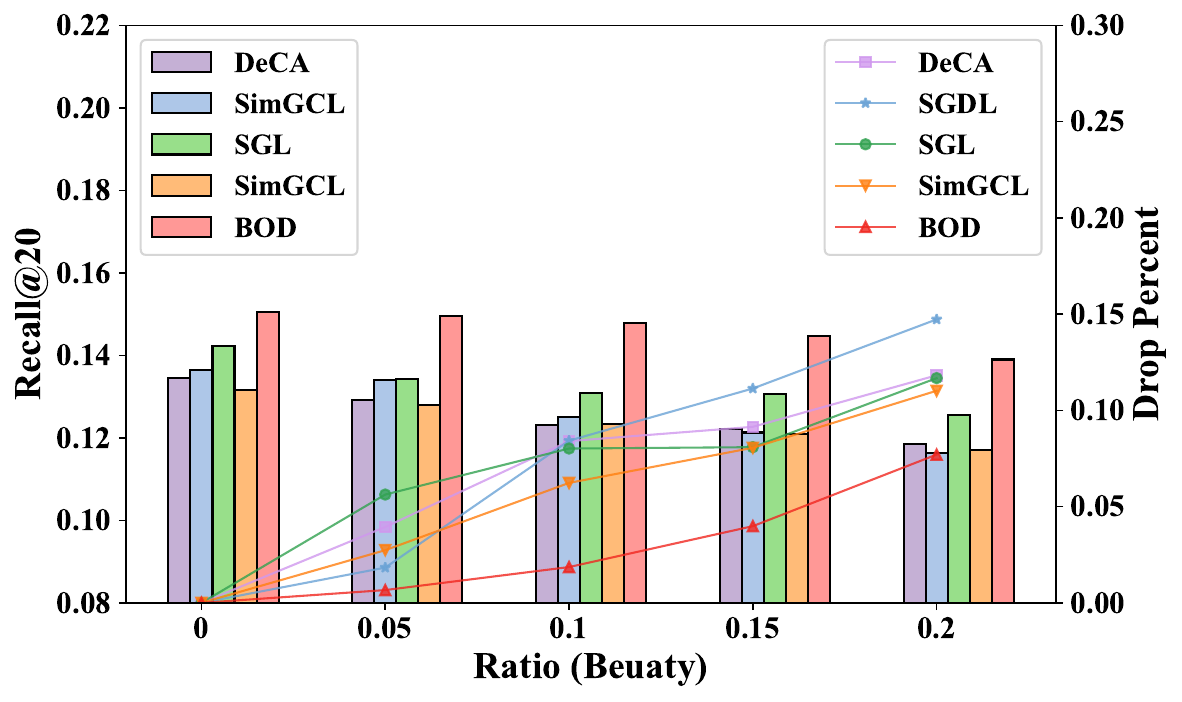}
  \caption{Model performance w.r.t. noise ratio.}
\label{robust comparison}
\end{figure}

\stitle{Robustness Comparison.} We also conduct experiments to check $\mathsf{BOD}$'s robustness. Following previous work \cite{13wu2021self}, we add a certain proportion of adversarial examples (i.e., 5\%, 10\%, 15\%, 20\% negative user-item interactions) into the training set, while keeping the testing set unchanged. Figure.~\ref{robust comparison} shows the experimental results on Beauty dataset. We can see the results of $\mathsf{BOD}$ maintain the best in all cases, despite performances reduced as noise data adding. Besides, the trend of $\mathsf{BOD}$'s drop percent curve keeps the minor change, which illustrates that $\mathsf{BOD}$ is the least affected by noise. This suggests that the denoising process of $\mathsf{BOD}$ can better figure out useful patterns under noisy conditions.

\subsection{Time Complexity Analysis (RQ2)}
In this part, we set LightGCN as the base model and report the real running time of compared methods for one epoch. The results in Table ~\ref{runtime} are collected on an Intel(R) Core(TM) i9-10900X CPU and a GeForce RTX 3090Ti GPU. As shown in Table ~\ref{runtime}, we can see the running time increases with the volume of the datasets. Besides, the running speed of $\mathsf{BOD}$ is much quicker than denoising methods (T-CE, DeCA, and SGDL), even though they only deal with positive samples. Furthermore, SGL and SimGCL do not need additional time on denoising training, but the time cost of $\mathsf{BOD}$ is still less than those of SGL and SimGCL, which proves that the generator-based one-step gradient matching scheme can save much time from the computation.

\subsection{The Effect of Inner Optimization on BOD (RQ3)}
\label{Study of BOD}

In Equation 10 of Section 4.2, we design the objective function of the inner optimization using the weighted sum of the two recommendation losses.
However, in order to solve for the outer optimization, we only need the inner optimization to provide (two gradients).
It may appear that solving the inner optimization with the weighted sum of losses is unnecessary.
Yet, our analysis demonstrates the importance of including both losses.
For this purpose, we discuss the following three cases:

\begin{itemize}[leftmargin=12pt]
    \item $\mathsf{BOD}_{\text{BPR}}$: We calculate BPR and AU loss both to get corresponding gradients, but only use the BPR's gradient to optimize the recommendation model in the inner optimization.
    
    \item $\mathsf{BOD}_{\text{AU}}$: Similar to the previous case, the difference is that we use the gradient generated from AU loss to optimize the recommendation model.
    
    \item $\mathsf{BOD}_{\text{BPR+AU}}$: In this (default) case, we use BPR's gradient and AU's gradients with the weight $\alpha$ to control the balance, to optimize recommendation model.
\end{itemize}

It is worth reminding that we hardly execute $\mathsf{BOD}$ without the generator due to the expensive space cost ("out of memory" will be displayed). 
As Table~\ref{study of BOD} shows, we find that if we only use one loss to optimize the recommendation model in the inner optimization,
its performance is not as good as the case when two losses are used. 

\begin{table}[h]
\caption{Running time in seconds per epoch.} \vspace{-1em}
\resizebox{!}{.1\textwidth}{
\begin{tabular}{@{}c|c|c|c@{}}
\toprule
\textbf{method} & \textbf{Beauty} & \textbf{iFashion} & \textbf{Yelp2018} \\ \midrule
Normal          & 3.47(+0.08)            & 89.22(+3.86)                     & 63.39(+3.23)             \\ \midrule
T-CE            & 41.5(+2.06)            & 765.64(+26.87)                    & 631.45(+21.33)            \\
DeCA            & 25.10(+1.28)           & 556.50(+18.78)                    & 478.68(+19.01)            \\
SGDL            & 32.12(+1.82)           & 688.21(+20.43)                    & 564.46(+16.92)            \\
SGL             & 8.12(+0.58)            & 302.32(+12.26)                    & 270.17(+11.81)            \\
SimGCL          & 7.26(+0.40)            & 273.66(+11.31)                    & 159.24(+8.01)            \\ \midrule
$\mathsf{BOD}$             & \textbf{6.82(+0.16)}   & \textbf{97.38(+4.11)}            & \textbf{75.98(+3.44)}    \\ \bottomrule
\end{tabular}}
\label{runtime}
\end{table}

\begin{table}[h]
\caption{The Effect of Inner optimization on $\mathsf{BOD}$.}
\resizebox{.48\textwidth}{!}{
\begin{tabular}{@{}cc|cc|cc|cc@{}}
\toprule
\multicolumn{2}{c|}{\textbf{Dataset}}                            & \multicolumn{2}{c|}{\textbf{Beauty}}                   & \multicolumn{2}{c|}{\textbf{iFashion}}         & \multicolumn{2}{c}{\textbf{Yelp2018}}                  \\ \midrule
\multicolumn{1}{c|}{Base Model}                & Component       & \multicolumn{1}{c|}{R@20}            & N@20            & \multicolumn{1}{c|}{R@20}            & N@20            & \multicolumn{1}{c|}{R@20}            & N@20            \\ \midrule
\multicolumn{1}{c|}{\multirow{3}{*}{GMF}}      & $\mathsf{BOD}_{\text{BPR}}$        & \multicolumn{1}{c|}{0.1029}          & 0.0599          & \multicolumn{1}{c|}{0.0664}          & 0.0306          & \multicolumn{1}{c|}{0.0531}          & 0.0422          \\
\multicolumn{1}{c|}{}                          & $\mathsf{BOD}_{\text{AU}}$ & \multicolumn{1}{c|}{0.1133}          & 0.0659          & \multicolumn{1}{c|}{0.0705}          & 0.0313          & \multicolumn{1}{c|}{0.0547}          & 0.0461          \\
\multicolumn{1}{c|}{}                          & $\mathsf{BOD}_{\text{BPR+AU}}$             & \multicolumn{1}{c|}{\textbf{0.1193}} & \textbf{0.0688} & \multicolumn{1}{c|}{\textbf{0.0860}} & \textbf{0.0426} & \multicolumn{1}{c|}{\textbf{0.0621}} & \textbf{0.0513} \\ \midrule
\multicolumn{1}{c|}{\multirow{3}{*}{NCF}}      & $\mathsf{BOD}_{\text{BPR}}$        & \multicolumn{1}{c|}{0.1208}          & 0.0683          & \multicolumn{1}{c|}{0.0806}          & 0.0385          & \multicolumn{1}{c|}{0.0564}          & 0.0432          \\
\multicolumn{1}{c|}{}                          & $\mathsf{BOD}_{\text{AU}}$ & \multicolumn{1}{c|}{0.1289}          & 0.0739          & \multicolumn{1}{c|}{0.0824}          & 0.0386          & \multicolumn{1}{c|}{0.0511}          & 0.0411          \\
\multicolumn{1}{c|}{}                          & $\mathsf{BOD}_{\text{BPR+AU}}$             & \multicolumn{1}{c|}{\textbf{0.1352}} & \textbf{0.0774} & \multicolumn{1}{c|}{\textbf{0.0909}} & \textbf{0.0442} & \multicolumn{1}{c|}{\textbf{0.0633}} & \textbf{0.0531} \\ \midrule
\multicolumn{1}{c|}{\multirow{3}{*}{NGCF}}     & $\mathsf{BOD}_{\text{BPR}}$        & \multicolumn{1}{c|}{0.1295}          & 0.0715          & \multicolumn{1}{c|}{0.0976}          & 0.0354          & \multicolumn{1}{c|}{0.0476}          & 0.0406          \\
\multicolumn{1}{c|}{}                          & $\mathsf{BOD}_{\text{AU}}$ & \multicolumn{1}{c|}{0.1297}          & 0.0717          & \multicolumn{1}{c|}{0.0983}          & 0.0385          & \multicolumn{1}{c|}{0.0545}          & 0.0454          \\
\multicolumn{1}{c|}{}                          & $\mathsf{BOD}_{\text{AU}}$             & \multicolumn{1}{c|}{\textbf{0.1415}} & \textbf{0.0827} & \multicolumn{1}{c|}{\textbf{0.1053}} & \textbf{0.0480} & \multicolumn{1}{c|}{\textbf{0.0565}} & \textbf{0.0466} \\ \midrule
\multicolumn{1}{c|}{\multirow{3}{*}{LightGCN}} & $\mathsf{BOD}_{\text{BPR}}$        & \multicolumn{1}{c|}{0.1386}          & 0.0868          & \multicolumn{1}{c|}{0.0984}          & 0.0415          & \multicolumn{1}{c|}{0.0598}          & 0.0523          \\
\multicolumn{1}{c|}{}                          & $\mathsf{BOD}_{\text{AU}}$ & \multicolumn{1}{c|}{0.1400}          & 0.0876          & \multicolumn{1}{c|}{0.1054}          & 0.0443          & \multicolumn{1}{c|}{0.0625}          & 0.0546          \\
\multicolumn{1}{c|}{}                          & $\mathsf{BOD}_{\text{BPR+AU}}$             & \multicolumn{1}{c|}{\textbf{0.1548}} & \textbf{0.0895} & \multicolumn{1}{c|}{\textbf{0.1180}} & \textbf{0.0576} & \multicolumn{1}{c|}{\textbf{0.7000}} & \textbf{0.0588} \\ \bottomrule
\end{tabular}}
\label{study of BOD}
\end{table}

\subsection{The Effect of Generator on BOD (RQ4)}
We delve deeper into the impact of generator design on performance. For this reason, we compare the applied AE method (using 2-layer Multi-Layer Perceptron, denoted as 2-MLP) with the classical variational autoencoder method (denoted as VAE) \cite{64kingma2013auto}. We also change the number of MLP layers to 1 and 3 (denoted as 1-MLP and 3-MLP) to further illustrate the difference. We perform experiments on the Beauty dataset and use LightGCN as the recommendation encoder. The experimental in Table~\ref{Generator Study} outcomes demonstrate that employing the 2-MLP method as a generator can yield improvements to a certain extent compared to the classical VAE method. In addition, the 2-MLP increases with the number of MLP layers and reaches stability at 2 layers. Therefore, we used the 2-layer MLP based AE method in the paper.

\begin{table}[h]
\caption{The Effect of Generator on $\mathsf{BOD}$.}\vspace{-1em}
\resizebox{.44\textwidth}{!}{
\begin{tabular}{@{}c|c|c|c|c@{}}
\toprule
\textbf{Method} & \textbf{Recall@10} & \textbf{Recall@20} & \textbf{NDCG@10} & \textbf{NDCG@20} \\ \midrule
1-MLP           & 0.1011             & 0.1498             & 0.0718           & 0.0855           \\
3-MLP           & 0.1095             & 0.1548             & 0.0755           & 0.0894           \\
VAE             & 0.1074             & 0.1532             & 0.0731           & 0.0888           \\ \midrule
2-MLP (ours)     & \textbf{0.1095}    & \textbf{0.1548}    & \textbf{0.0755}  & \textbf{0.0895}  \\ \bottomrule
\end{tabular}}
\label{Generator Study}
\end{table}

\subsection{Parameters Sensitivity Analysis (RQ5)}
\label{section_parameters}
\begin{figure}[h]
  \centering
  \includegraphics[width=\linewidth]{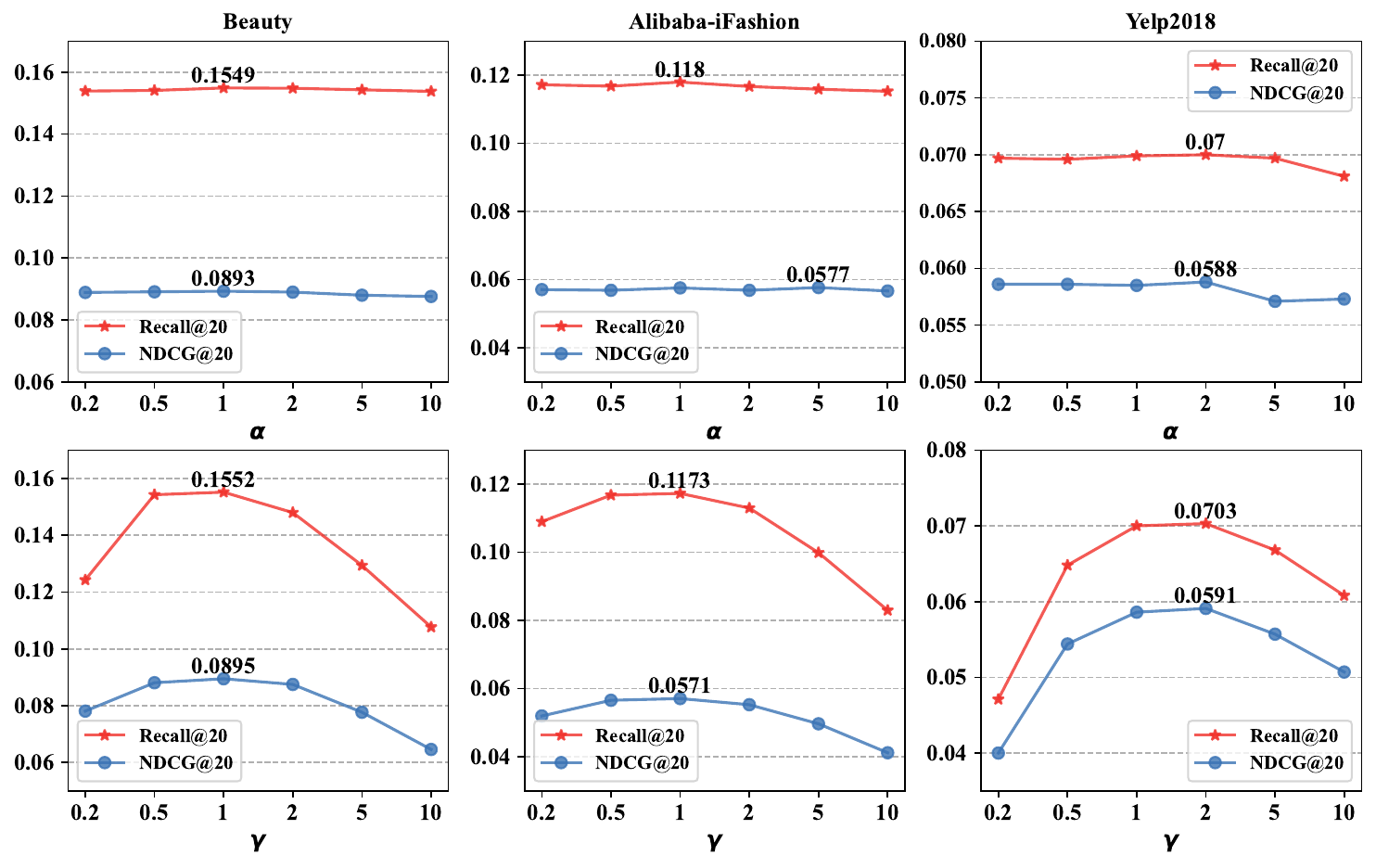}
  \caption{Parameter sensitivity with regard to $\alpha$ and $\gamma$.} 
\label{Parameters Sensitivity} 
\end{figure}

We investigate the model’s sensitivity to the hyperparameters $\alpha$ in Equation \ref{equation8} and $\gamma$ in Equation \ref{equation11} (see Appendix). We fix $\alpha$ as 1 and try different values of $\gamma$ with the set [0.2, 0.5, 1, 2, 5, 10], then we fix $\gamma$ as 1 and try different values of $\alpha$ with the set [0.2, 0.5, 1, 2, 5, 10]. As shown in Fig.~\ref{Parameters Sensitivity}, we can observe the performance keep stable as $\alpha$ changes. Besides, a similar trend in the performance of $\gamma$ increases first and then decreases on all the datasets. $\mathsf{BOD}$ reaches its best performance when $\gamma$ is in [0.5, 1, 2], which suggests that the weights of alignment and uniformity should be balanced at the same level.

\subsection{Generalizability Analysis (RQ6)}
We emphasize that our framework $\mathsf{BOD}$ is generic, which means that it can apply to any recommendation model in a plug-and-play fashion.
One thing of interest is whether we can further improve the performance of the latest robust recommendation models SGL and SimGCL. 
To this end, we use SGL and SimGCL as base models for $\mathsf{BOD}$, and the results are shown in Fig.~5.
From the experimental results, we conclude that $\mathsf{BOD}$ does further improve the performance of these two methods.
This confirms the generalizability of our method.

\begin{figure}[h]
  \centering
  \includegraphics[width=\linewidth]{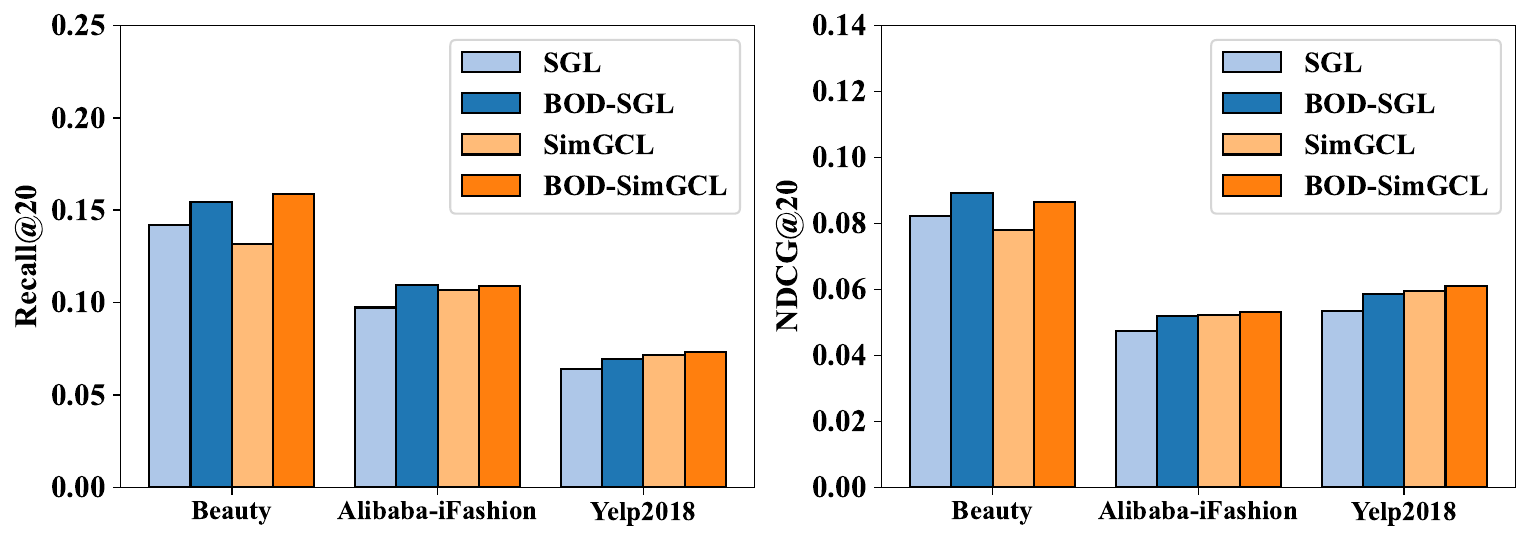}
  \caption{The generalizability of $\mathsf{BOD}$ to SGL and SimGCL.}
\label{transferity}
\end{figure}

\section{RELATED WORK}
Our study endeavors to denoise recommendations by employing a bi-level optimization framework. To begin with, we present various bi-level optimization methods applied in recommendation. Concurrently, researchers attempt to denoise implicit feedback through many avenues, and we mainly focus on denoising methods without extra information in this paper.

\stitle{Bi-level Optimization in Recommendation.} Bi-level optimization techniques have been extensively utilized in the domain of recommender systems \cite{61he2018adversarial, 62chae2018cfgan, 63ganhor2022unlearning}. For instance, APR introduces adversarial perturbations to model parameters as a means to improve recommendations through adversarial training. IRGAN employs a game-theoretical minimax game to iteratively optimize both a generative model and a discriminative model, ultimately generating refined recommendation results. Adv-MultVAE incorporates adversarial training into MultVAE architecture to eliminate implicit information pertaining to protected attributes while preserving recommendation performance. These approaches showcase the effectiveness of bi-level optimization in enhancing recommendation performance, yet there is currently a dearth of research addressing denoising within this context.

\stitle{Denoising Methods.} 
A straightforward way approach \cite{08gantner2012personalized,54wang2021implicit} to denoising implicit feedback is to select clean and informative samples and use them to train a recommendation model. For example, WBPR \cite{08gantner2012personalized} notices that popular items are more likely real negative instances if they have missing actions, and then samples popular items with higher probabilities. IR \cite{54wang2021implicit} produces synthetic labels for user preferences depending on the distinction between labels and predictions, to uncover false-positive and false-negative samples. However, the variability of their performance is wide due to their reliance on the sampling distribution \cite{07yuan2018fbgd}. Thus, some methods \cite{10wang2022learning,11wang2021denoising,12DBLP:conf/sigir/Gao0HCZFZ22} provide contribution weight for each sample during recommendation model training. T-CE \cite{11wang2021denoising} proposes to assign lower weights to large loss samples with a dynamic threshold because they find that a sample with high loss is more likely to be noisy. DeCA \cite{10wang2022learning} assumes that different models make relatively similar predictions on clean examples, so it incorporates the training process of two recommendation models and distinguishes clean examples from positive samples based on two different predictions. SGDL \cite{12DBLP:conf/sigir/Gao0HCZFZ22} claims that models tend to dig out clean samples in the initial training stages, and then adaptively assign weights for samples based on the similarity of collected clean samples. We can notice that existing methods often resort to prior
knowledge for determining weights, but prior knowledge has its own scope of application since we cannot be sure of the validity of prior knowledge itself.

\stitle{Other Directions.} There are also some robust recommendation methods \cite{59yu2022self} that do not belong to the above methods. For example, SGL \cite{13wu2021self} creates two additional views of graphs by adding and dropping edges and leverages contrastive learning to distinguish hard noisy samples. Recently, several works have considered augmentation from the perspective of embeddings. For example, SimGCL \cite{15yu2022graph} considers embeddings distribution, and thus trains models to force the distribution of embeddings to fit uniformity. 

It should be noted that previous research often uses additional information (attribute features \cite{44zhang2022neuro}, social information \cite{02yu2021self}, etc.) to process weights. This paper discusses denoising tasks in extreme situations where only implicit feedback information can be obtained, so we do not introduce some denoising methods with extra information \cite{03zhang2019nonlinear,05ding2019sampler,23lian2020geography,36tang2013exploiting, 58yin2015joint}.

\section{CONCLUSION}
In this paper, we model recommendation denoising as bi-level optimization and propose an efficient solution for the proposed bi-level optimization framework, called $\mathsf{BOD}$. In $\mathsf{BOD}$, a generator-based parameter generation technique to save space, and a gradient-matching-based parameter-solving technique to save time. Extensive experiments on three real-world datasets demonstrate the superiority of $\mathsf{BOD}$. 
Moving forward, our future research will focus on investigating more efficient approaches within the bi-level optimization framework to address specific challenges associated with recommendation denoising. One existing challenge pertains to the presence of noisy data in the test sets, which cannot be effectively identified during the bi-level optimization process. 


\begin{acks}
This work is partially supported by the National Natural Science Foundation of China (62176028), Australian Research Council Future Fellowship (Grant No. FT210100624) and Discovery Project (Grant No. DP190101985).
\end{acks}

\balance
\bibliographystyle{ACM-Reference-Format}
\bibliography{BOD}


\begin{thebibliography}{49}


\ifx \showCODEN    \undefined \def \showCODEN     #1{\unskip}     \fi
\ifx \showDOI      \undefined \def \showDOI       #1{#1}\fi
\ifx \showISBNx    \undefined \def \showISBNx     #1{\unskip}     \fi
\ifx \showISBNxiii \undefined \def \showISBNxiii  #1{\unskip}     \fi
\ifx \showISSN     \undefined \def \showISSN      #1{\unskip}     \fi
\ifx \showLCCN     \undefined \def \showLCCN      #1{\unskip}     \fi
\ifx \shownote     \undefined \def \shownote      #1{#1}          \fi
\ifx \showarticletitle \undefined \def \showarticletitle #1{#1}   \fi
\ifx \showURL      \undefined \def \showURL       {\relax}        \fi
\providecommand\bibfield[2]{#2}
\providecommand\bibinfo[2]{#2}
\providecommand\natexlab[1]{#1}
\providecommand\showeprint[2][]{arXiv:#2}

\bibitem[Bian et~al\mbox{.}(2021)]%
        {27bian2021denoising}
\bibfield{author}{\bibinfo{person}{Zhi Bian}, \bibinfo{person}{Shaojun Zhou},
  \bibinfo{person}{Hao Fu}, \bibinfo{person}{Qihong Yang},
  \bibinfo{person}{Zhenqi Sun}, \bibinfo{person}{Junjie Tang},
  \bibinfo{person}{Guiquan Liu}, \bibinfo{person}{Kaikui Liu}, {and}
  \bibinfo{person}{Xiaolong Li}.} \bibinfo{year}{2021}\natexlab{}.
\newblock \showarticletitle{Denoising user-aware memory network for
  recommendation}. In \bibinfo{booktitle}{\emph{Proceedings of ACM Conference
  on Recommender Systems 2021}}. \bibinfo{pages}{400--410}.
\newblock


\bibitem[Chae et~al\mbox{.}(2018)]%
        {62chae2018cfgan}
\bibfield{author}{\bibinfo{person}{Dong-Kyu Chae}, \bibinfo{person}{Jin-Soo
  Kang}, \bibinfo{person}{Sang-Wook Kim}, {and} \bibinfo{person}{Jung-Tae
  Lee}.} \bibinfo{year}{2018}\natexlab{}.
\newblock \showarticletitle{Cfgan: A generic collaborative filtering framework
  based on generative adversarial networks}. In
  \bibinfo{booktitle}{\emph{Proceedings of the 27th ACM international
  conference on information and knowledge management}}.
  \bibinfo{pages}{137--146}.
\newblock


\bibitem[Chen et~al\mbox{.}(2021)]%
        {47chen2021autodebias}
\bibfield{author}{\bibinfo{person}{Jiawei Chen}, \bibinfo{person}{Hande Dong},
  \bibinfo{person}{Yang Qiu}, \bibinfo{person}{Xiangnan He},
  \bibinfo{person}{Xin Xin}, \bibinfo{person}{Liang Chen},
  \bibinfo{person}{Guli Lin}, {and} \bibinfo{person}{Keping Yang}.}
  \bibinfo{year}{2021}\natexlab{}.
\newblock \showarticletitle{AutoDebias: Learning to debias for recommendation}.
  In \bibinfo{booktitle}{\emph{Proceedings of the International ACM SIGIR
  Conference on Research and Development in Information Retrieval 2021}}.
  \bibinfo{pages}{21--30}.
\newblock


\bibitem[Ding et~al\mbox{.}(2019)]%
        {05ding2019sampler}
\bibfield{author}{\bibinfo{person}{Jingtao Ding}, \bibinfo{person}{Guanghui
  Yu}, \bibinfo{person}{Xiangnan He}, \bibinfo{person}{Fuli Feng},
  \bibinfo{person}{Yong Li}, {and} \bibinfo{person}{Depeng Jin}.}
  \bibinfo{year}{2019}\natexlab{}.
\newblock \showarticletitle{Sampler design for bayesian personalized ranking by
  leveraging view data}.
\newblock \bibinfo{journal}{\emph{IEEE Transactions on Knowledge and Data
  Engineering}} \bibinfo{volume}{33}, \bibinfo{number}{2}
  (\bibinfo{year}{2019}), \bibinfo{pages}{667--681}.
\newblock


\bibitem[Ganh{\"o}r et~al\mbox{.}(2022)]%
        {63ganhor2022unlearning}
\bibfield{author}{\bibinfo{person}{Christian Ganh{\"o}r},
  \bibinfo{person}{David Penz}, \bibinfo{person}{Navid Rekabsaz},
  \bibinfo{person}{Oleg Lesota}, {and} \bibinfo{person}{Markus Schedl}.}
  \bibinfo{year}{2022}\natexlab{}.
\newblock \showarticletitle{Unlearning Protected User Attributes in
  Recommendations with Adversarial Training}. In
  \bibinfo{booktitle}{\emph{Proceedings of the 45th International ACM SIGIR
  Conference on Research and Development in Information Retrieval}}.
  \bibinfo{pages}{2142--2147}.
\newblock


\bibitem[Gantner et~al\mbox{.}(2012)]%
        {08gantner2012personalized}
\bibfield{author}{\bibinfo{person}{Zeno Gantner}, \bibinfo{person}{Lucas
  Drumond}, \bibinfo{person}{Christoph Freudenthaler}, {and}
  \bibinfo{person}{Lars Schmidt-Thieme}.} \bibinfo{year}{2012}\natexlab{}.
\newblock \showarticletitle{Personalized ranking for non-uniformly sampled
  items}. In \bibinfo{booktitle}{\emph{Proceedings of KDD Cup 2011}}. PMLR,
  \bibinfo{pages}{231--247}.
\newblock


\bibitem[Gao et~al\mbox{.}(2022)]%
        {12DBLP:conf/sigir/Gao0HCZFZ22}
\bibfield{author}{\bibinfo{person}{Yunjun Gao}, \bibinfo{person}{Yuntao Du},
  \bibinfo{person}{Yujia Hu}, \bibinfo{person}{Lu Chen},
  \bibinfo{person}{Xinjun Zhu}, \bibinfo{person}{Ziquan Fang}, {and}
  \bibinfo{person}{Baihua Zheng}.} \bibinfo{year}{2022}\natexlab{}.
\newblock \showarticletitle{Self-Guided Learning to Denoise for Robust
  Recommendation}. In \bibinfo{booktitle}{\emph{Proceedings of the
  International ACM SIGIR Conference on Research and Development in Information
  Retrieval 2022}}. \bibinfo{publisher}{{ACM}}, \bibinfo{pages}{1412--1422}.
\newblock


\bibitem[Glorot and Bengio(2010)]%
        {42glorot2010understanding}
\bibfield{author}{\bibinfo{person}{Xavier Glorot} {and} \bibinfo{person}{Yoshua
  Bengio}.} \bibinfo{year}{2010}\natexlab{}.
\newblock \showarticletitle{Understanding the difficulty of training deep
  feedforward neural networks}. In \bibinfo{booktitle}{\emph{Proceedings of the
  thirteenth international conference on artificial intelligence and statistics
  2010}}. JMLR, \bibinfo{pages}{249--256}.
\newblock


\bibitem[Guo et~al\mbox{.}(2019)]%
        {55guo2019streaming}
\bibfield{author}{\bibinfo{person}{Lei Guo}, \bibinfo{person}{Hongzhi Yin},
  \bibinfo{person}{Qinyong Wang}, \bibinfo{person}{Tong Chen},
  \bibinfo{person}{Alexander Zhou}, {and} \bibinfo{person}{Nguyen Quoc
  Viet~Hung}.} \bibinfo{year}{2019}\natexlab{}.
\newblock \showarticletitle{Streaming session-based recommendation}. In
  \bibinfo{booktitle}{\emph{Proceedings of the ACM SIGKDD Conference on
  Knowledge Discovery and Data Mining 2019}}. \bibinfo{pages}{1569--1577}.
\newblock


\bibitem[He and Chua(2017)]%
        {39he2017neural}
\bibfield{author}{\bibinfo{person}{Xiangnan He} {and} \bibinfo{person}{Tat-Seng
  Chua}.} \bibinfo{year}{2017}\natexlab{}.
\newblock \showarticletitle{Neural factorization machines for sparse predictive
  analytics}. In \bibinfo{booktitle}{\emph{Proceedings of the International ACM
  SIGIR Conference on Research and Development in Information Retrieval 2017}}.
  \bibinfo{pages}{355--364}.
\newblock


\bibitem[He et~al\mbox{.}(2020)]%
        {26he2020lightgcn}
\bibfield{author}{\bibinfo{person}{Xiangnan He}, \bibinfo{person}{Kuan Deng},
  \bibinfo{person}{Xiang Wang}, \bibinfo{person}{Yan Li},
  \bibinfo{person}{Yongdong Zhang}, {and} \bibinfo{person}{Meng Wang}.}
  \bibinfo{year}{2020}\natexlab{}.
\newblock \showarticletitle{Lightgcn: Simplifying and powering graph
  convolution network for recommendation}. In
  \bibinfo{booktitle}{\emph{Proceedings of the International ACM SIGIR
  Conference on Research and Development in Information Retrieval 2020}}.
  \bibinfo{pages}{639--648}.
\newblock


\bibitem[He et~al\mbox{.}(2018)]%
        {61he2018adversarial}
\bibfield{author}{\bibinfo{person}{Xiangnan He}, \bibinfo{person}{Zhankui He},
  \bibinfo{person}{Xiaoyu Du}, {and} \bibinfo{person}{Tat-Seng Chua}.}
  \bibinfo{year}{2018}\natexlab{}.
\newblock \showarticletitle{Adversarial personalized ranking for
  recommendation}. In \bibinfo{booktitle}{\emph{The 41st International ACM
  SIGIR conference on research \& development in information retrieval}}.
  \bibinfo{pages}{355--364}.
\newblock


\bibitem[He et~al\mbox{.}(2017)]%
        {50he2017neural}
\bibfield{author}{\bibinfo{person}{Xiangnan He}, \bibinfo{person}{Lizi Liao},
  \bibinfo{person}{Hanwang Zhang}, \bibinfo{person}{Liqiang Nie},
  \bibinfo{person}{Xia Hu}, {and} \bibinfo{person}{Tat-Seng Chua}.}
  \bibinfo{year}{2017}\natexlab{}.
\newblock \showarticletitle{Neural collaborative filtering}. In
  \bibinfo{booktitle}{\emph{Proceedings of the 26th international conference on
  world wide web}}. \bibinfo{pages}{173--182}.
\newblock


\bibitem[Hu et~al\mbox{.}(2008)]%
        {60hu2008collaborative}
\bibfield{author}{\bibinfo{person}{Yifan Hu}, \bibinfo{person}{Yehuda Koren},
  {and} \bibinfo{person}{Chris Volinsky}.} \bibinfo{year}{2008}\natexlab{}.
\newblock \showarticletitle{Collaborative filtering for implicit feedback
  datasets}. In \bibinfo{booktitle}{\emph{IEEE International Conference on Data
  Mining 2008}}. \bibinfo{pages}{263--272}.
\newblock


\bibitem[Jin et~al\mbox{.}(2022)]%
        {53jin2022condensing}
\bibfield{author}{\bibinfo{person}{Wei Jin}, \bibinfo{person}{Xianfeng Tang},
  \bibinfo{person}{Haoming Jiang}, \bibinfo{person}{Zheng Li},
  \bibinfo{person}{Danqing Zhang}, \bibinfo{person}{Jiliang Tang}, {and}
  \bibinfo{person}{Bing Yin}.} \bibinfo{year}{2022}\natexlab{}.
\newblock \showarticletitle{Condensing graphs via one-step gradient matching}.
  In \bibinfo{booktitle}{\emph{Proceedings of the ACM SIGKDD Conference on
  Knowledge Discovery and Data Mining 2022}}. \bibinfo{pages}{720--730}.
\newblock


\bibitem[Jin et~al\mbox{.}(2021)]%
        {19jin2021graph}
\bibfield{author}{\bibinfo{person}{Wei Jin}, \bibinfo{person}{Lingxiao Zhao},
  \bibinfo{person}{Shichang Zhang}, \bibinfo{person}{Yozen Liu},
  \bibinfo{person}{Jiliang Tang}, {and} \bibinfo{person}{Neil Shah}.}
  \bibinfo{year}{2021}\natexlab{}.
\newblock \showarticletitle{Graph Condensation for Graph Neural Networks}. In
  \bibinfo{booktitle}{\emph{International Conference on Learning
  Representations 2021}}.
\newblock


\bibitem[Kingma and Ba(2014)]%
        {41kingma2014adam}
\bibfield{author}{\bibinfo{person}{Diederik~P Kingma} {and}
  \bibinfo{person}{Jimmy Ba}.} \bibinfo{year}{2014}\natexlab{}.
\newblock \showarticletitle{Adam: A method for stochastic optimization}.
\newblock \bibinfo{journal}{\emph{arXiv preprint arXiv:1412.6980}}
  (\bibinfo{year}{2014}).
\newblock


\bibitem[Kingma and Welling(2013)]%
        {64kingma2013auto}
\bibfield{author}{\bibinfo{person}{Diederik~P Kingma} {and}
  \bibinfo{person}{Max Welling}.} \bibinfo{year}{2013}\natexlab{}.
\newblock \showarticletitle{Auto-encoding variational bayes}.
\newblock \bibinfo{journal}{\emph{arXiv preprint arXiv:1312.6114}}
  (\bibinfo{year}{2013}).
\newblock


\bibitem[Lian et~al\mbox{.}(2020)]%
        {23lian2020geography}
\bibfield{author}{\bibinfo{person}{Defu Lian}, \bibinfo{person}{Yongji Wu},
  \bibinfo{person}{Yong Ge}, \bibinfo{person}{Xing Xie}, {and}
  \bibinfo{person}{Enhong Chen}.} \bibinfo{year}{2020}\natexlab{}.
\newblock \showarticletitle{Geography-aware sequential location
  recommendation}. In \bibinfo{booktitle}{\emph{Proceedings of the ACM SIGKDD
  International Conference on Knowledge Discovery \& Data Mining 2020}}.
  \bibinfo{pages}{2009--2019}.
\newblock


\bibitem[Lu et~al\mbox{.}(2018)]%
        {48lu2018between}
\bibfield{author}{\bibinfo{person}{Hongyu Lu}, \bibinfo{person}{Min Zhang},
  {and} \bibinfo{person}{Shaoping Ma}.} \bibinfo{year}{2018}\natexlab{}.
\newblock \showarticletitle{Between clicks and satisfaction: Study on
  multi-phase user preferences and satisfaction for online news reading}. In
  \bibinfo{booktitle}{\emph{Proceedings of the International ACM SIGIR
  Conference on Research and Development in Information Retrieval 2018}}.
  \bibinfo{pages}{435--444}.
\newblock


\bibitem[Nguyen et~al\mbox{.}(2021)]%
        {04nguyen2021dataset}
\bibfield{author}{\bibinfo{person}{Timothy Nguyen}, \bibinfo{person}{Roman
  Novak}, \bibinfo{person}{Lechao Xiao}, {and} \bibinfo{person}{Jaehoon Lee}.}
  \bibinfo{year}{2021}\natexlab{}.
\newblock \showarticletitle{Dataset distillation with infinitely wide
  convolutional networks}.
\newblock \bibinfo{journal}{\emph{Advances in Neural Information Processing
  Systems}}  \bibinfo{volume}{34} (\bibinfo{year}{2021}),
  \bibinfo{pages}{5186--5198}.
\newblock


\bibitem[Pan and Chen(2013)]%
        {28pan2013gbpr}
\bibfield{author}{\bibinfo{person}{Weike Pan} {and} \bibinfo{person}{Li Chen}.}
  \bibinfo{year}{2013}\natexlab{}.
\newblock \showarticletitle{Gbpr: Group preference based bayesian personalized
  ranking for one-class collaborative filtering}. In
  \bibinfo{booktitle}{\emph{Proceedings of International Joint Conference on
  Artificial Intelligence 2013}}.
\newblock


\bibitem[Rendle et~al\mbox{.}(2014)]%
        {20rendle2014bayesian}
\bibfield{author}{\bibinfo{person}{Steffen Rendle}, \bibinfo{person}{Christoph
  Freudenthaler}, \bibinfo{person}{Zeno Gantner}, {and}
  \bibinfo{person}{Lars~BPR Schmidt-Thieme}.} \bibinfo{year}{2014}\natexlab{}.
\newblock \showarticletitle{Bayesian personalized ranking from implicit
  feedback}. In \bibinfo{booktitle}{\emph{Proc. of Uncertainty in Artificial
  Intelligence 2014}}. \bibinfo{pages}{452--461}.
\newblock


\bibitem[Rifai et~al\mbox{.}(2011)]%
        {45rifai2011contractive}
\bibfield{author}{\bibinfo{person}{Salah Rifai}, \bibinfo{person}{Pascal
  Vincent}, \bibinfo{person}{Xavier Muller}, \bibinfo{person}{Xavier Glorot},
  {and} \bibinfo{person}{Yoshua Bengio}.} \bibinfo{year}{2011}\natexlab{}.
\newblock \showarticletitle{Contractive auto-encoders: Explicit invariance
  during feature extraction}. In \bibinfo{booktitle}{\emph{Proceedings of the
  28th international conference on international conference on machine learning
  2011}}. \bibinfo{pages}{833--840}.
\newblock


\bibitem[Tang et~al\mbox{.}(2013)]%
        {36tang2013exploiting}
\bibfield{author}{\bibinfo{person}{Jiliang Tang}, \bibinfo{person}{Xia Hu},
  \bibinfo{person}{Huiji Gao}, {and} \bibinfo{person}{Huan Liu}.}
  \bibinfo{year}{2013}\natexlab{}.
\newblock \showarticletitle{Exploiting local and global social context for
  recommendation.}. In \bibinfo{booktitle}{\emph{Proceedings of International
  Joint Conference on Artificial Intelligence 2013}},
  Vol.~\bibinfo{volume}{13}. \bibinfo{pages}{2712--2718}.
\newblock


\bibitem[Wang et~al\mbox{.}(2022b)]%
        {17wang2022towards}
\bibfield{author}{\bibinfo{person}{Chenyang Wang}, \bibinfo{person}{Yuanqing
  Yu}, \bibinfo{person}{Weizhi Ma}, \bibinfo{person}{Min Zhang},
  \bibinfo{person}{Chong Chen}, \bibinfo{person}{Yiqun Liu}, {and}
  \bibinfo{person}{Shaoping Ma}.} \bibinfo{year}{2022}\natexlab{b}.
\newblock \showarticletitle{Towards Representation Alignment and Uniformity in
  Collaborative Filtering}. In \bibinfo{booktitle}{\emph{Proceedings of the ACM
  SIGKDD Conference on Knowledge Discovery and Data Mining 2022}}.
  \bibinfo{pages}{1816--1825}.
\newblock


\bibitem[Wang et~al\mbox{.}(2018)]%
        {24wang2018minimax}
\bibfield{author}{\bibinfo{person}{J Wang}, \bibinfo{person}{L Yu},
  \bibinfo{person}{W Zhang}, \bibinfo{person}{Y Gong}, \bibinfo{person}{Y Xu},
  \bibinfo{person}{B Wang}, \bibinfo{person}{P Zhang}, {and} \bibinfo{person}{D
  Zhang}.} \bibinfo{year}{2018}\natexlab{}.
\newblock \showarticletitle{A minimax game for unifying generative and
  discriminative information retrieval models}.
\newblock \bibinfo{journal}{\emph{Proceedings of the International ACM SIGIR
  Conference on Research and Development in Information Retrieval 2018}}
  (\bibinfo{year}{2018}).
\newblock


\bibitem[Wang et~al\mbox{.}(2019b)]%
        {56wang2019enhancing}
\bibfield{author}{\bibinfo{person}{Qinyong Wang}, \bibinfo{person}{Hongzhi
  Yin}, \bibinfo{person}{Hao Wang}, \bibinfo{person}{Quoc Viet~Hung Nguyen},
  \bibinfo{person}{Zi Huang}, {and} \bibinfo{person}{Lizhen Cui}.}
  \bibinfo{year}{2019}\natexlab{b}.
\newblock \showarticletitle{Enhancing collaborative filtering with generative
  augmentation}. In \bibinfo{booktitle}{\emph{Proceedings of the ACM SIGKDD
  Conference on Knowledge Discovery and Data Mining 2019}}.
  \bibinfo{pages}{548--556}.
\newblock


\bibitem[Wang and Isola(2020)]%
        {33wang2020understanding}
\bibfield{author}{\bibinfo{person}{Tongzhou Wang} {and}
  \bibinfo{person}{Phillip Isola}.} \bibinfo{year}{2020}\natexlab{}.
\newblock \showarticletitle{Understanding contrastive representation learning
  through alignment and uniformity on the hypersphere}. In
  \bibinfo{booktitle}{\emph{Proceedings of International Conference on Machine
  Learning 2020}}. PMLR, \bibinfo{pages}{9929--9939}.
\newblock


\bibitem[Wang et~al\mbox{.}(2021a)]%
        {11wang2021denoising}
\bibfield{author}{\bibinfo{person}{Wenjie Wang}, \bibinfo{person}{Fuli Feng},
  \bibinfo{person}{Xiangnan He}, \bibinfo{person}{Liqiang Nie}, {and}
  \bibinfo{person}{Tat-Seng Chua}.} \bibinfo{year}{2021}\natexlab{a}.
\newblock \showarticletitle{Denoising implicit feedback for recommendation}. In
  \bibinfo{booktitle}{\emph{Proceedings of the ACM International Conference on
  Web Search and Data Mining 2021}}. \bibinfo{pages}{373--381}.
\newblock


\bibitem[Wang et~al\mbox{.}(2019a)]%
        {51wang2019neural}
\bibfield{author}{\bibinfo{person}{Xiang Wang}, \bibinfo{person}{Xiangnan He},
  \bibinfo{person}{Meng Wang}, \bibinfo{person}{Fuli Feng}, {and}
  \bibinfo{person}{Tat-Seng Chua}.} \bibinfo{year}{2019}\natexlab{a}.
\newblock \showarticletitle{Neural graph collaborative filtering}. In
  \bibinfo{booktitle}{\emph{Proceedings of the 42nd international ACM SIGIR
  conference on Research and development in Information Retrieval}}.
  \bibinfo{pages}{165--174}.
\newblock


\bibitem[Wang et~al\mbox{.}(2022a)]%
        {10wang2022learning}
\bibfield{author}{\bibinfo{person}{Yu Wang}, \bibinfo{person}{Xin Xin},
  \bibinfo{person}{Zaiqiao Meng}, \bibinfo{person}{Joemon~M Jose},
  \bibinfo{person}{Fuli Feng}, {and} \bibinfo{person}{Xiangnan He}.}
  \bibinfo{year}{2022}\natexlab{a}.
\newblock \showarticletitle{Learning Robust Recommenders through Cross-Model
  Agreement}. In \bibinfo{booktitle}{\emph{Proceedings of the ACM Web
  Conference 2022}}. \bibinfo{pages}{2015--2025}.
\newblock


\bibitem[Wang et~al\mbox{.}(2021b)]%
        {30wang2021implicit}
\bibfield{author}{\bibinfo{person}{Zitai Wang}, \bibinfo{person}{Qianqian Xu},
  \bibinfo{person}{Zhiyong Yang}, \bibinfo{person}{Xiaochun Cao}, {and}
  \bibinfo{person}{Qingming Huang}.} \bibinfo{year}{2021}\natexlab{b}.
\newblock \showarticletitle{Implicit Feedbacks are Not Always Favorable:
  Iterative Relabeled One-Class Collaborative Filtering against Noisy
  Interactions}. In \bibinfo{booktitle}{\emph{Proceedings of the 29th ACM
  International Conference on Multimedia 2021}}. \bibinfo{pages}{3070--3078}.
\newblock


\bibitem[Wang et~al\mbox{.}(2021c)]%
        {54wang2021implicit}
\bibfield{author}{\bibinfo{person}{Zitai Wang}, \bibinfo{person}{Qianqian Xu},
  \bibinfo{person}{Zhiyong Yang}, \bibinfo{person}{Xiaochun Cao}, {and}
  \bibinfo{person}{Qingming Huang}.} \bibinfo{year}{2021}\natexlab{c}.
\newblock \showarticletitle{Implicit Feedbacks are Not Always Favorable:
  Iterative Relabeled One-Class Collaborative Filtering against Noisy
  Interactions}. In \bibinfo{booktitle}{\emph{Proceedings of the 29th ACM
  International Conference on Multimedia}}. \bibinfo{pages}{3070--3078}.
\newblock


\bibitem[Wu et~al\mbox{.}(2021a)]%
        {32wu2021ready}
\bibfield{author}{\bibinfo{person}{Fan Wu}, \bibinfo{person}{Min Gao},
  \bibinfo{person}{Junliang Yu}, \bibinfo{person}{Zongwei Wang},
  \bibinfo{person}{Kecheng Liu}, {and} \bibinfo{person}{Xu Wang}.}
  \bibinfo{year}{2021}\natexlab{a}.
\newblock \showarticletitle{Ready for emerging threats to recommender systems?
  A graph convolution-based generative shilling attack}.
\newblock \bibinfo{journal}{\emph{Information Sciences 2021}}
  \bibinfo{volume}{578} (\bibinfo{year}{2021}), \bibinfo{pages}{683--701}.
\newblock


\bibitem[Wu et~al\mbox{.}(2021b)]%
        {13wu2021self}
\bibfield{author}{\bibinfo{person}{Jiancan Wu}, \bibinfo{person}{Xiang Wang},
  \bibinfo{person}{Fuli Feng}, \bibinfo{person}{Xiangnan He},
  \bibinfo{person}{Liang Chen}, \bibinfo{person}{Jianxun Lian}, {and}
  \bibinfo{person}{Xing Xie}.} \bibinfo{year}{2021}\natexlab{b}.
\newblock \showarticletitle{Self-supervised graph learning for recommendation}.
  In \bibinfo{booktitle}{\emph{Proceedings of the International ACM SIGIR
  Conference on Research and Development in Information Retrieval 2021}}.
  \bibinfo{pages}{726--735}.
\newblock


\bibitem[Yin et~al\mbox{.}(2015)]%
        {58yin2015joint}
\bibfield{author}{\bibinfo{person}{Hongzhi Yin}, \bibinfo{person}{Bin Cui},
  \bibinfo{person}{Zi Huang}, \bibinfo{person}{Weiqing Wang},
  \bibinfo{person}{Xian Wu}, {and} \bibinfo{person}{Xiaofang Zhou}.}
  \bibinfo{year}{2015}\natexlab{}.
\newblock \showarticletitle{Joint modeling of users' interests and mobility
  patterns for point-of-interest recommendation}. In
  \bibinfo{booktitle}{\emph{Proceedings of the 23rd ACM International
  Conference on Multimedia}}. \bibinfo{pages}{819--822}.
\newblock


\bibitem[Yin et~al\mbox{.}(2019)]%
        {57yin2019social}
\bibfield{author}{\bibinfo{person}{Hongzhi Yin}, \bibinfo{person}{Qinyong
  Wang}, \bibinfo{person}{Kai Zheng}, \bibinfo{person}{Zhixu Li},
  \bibinfo{person}{Jiali Yang}, {and} \bibinfo{person}{Xiaofang Zhou}.}
  \bibinfo{year}{2019}\natexlab{}.
\newblock \showarticletitle{Social influence-based group representation
  learning for group recommendation}. In \bibinfo{booktitle}{\emph{IEEE 35th
  International Conference on Data Engineering 2019}}.
  \bibinfo{pages}{566--577}.
\newblock


\bibitem[Yu et~al\mbox{.}(2021a)]%
        {25yu2021socially}
\bibfield{author}{\bibinfo{person}{Junliang Yu}, \bibinfo{person}{Hongzhi Yin},
  \bibinfo{person}{Min Gao}, \bibinfo{person}{Xin Xia},
  \bibinfo{person}{Xiangliang Zhang}, {and} \bibinfo{person}{Nguyen~Quoc
  Viet~Hung}.} \bibinfo{year}{2021}\natexlab{a}.
\newblock \showarticletitle{Socially-aware self-supervised tri-training for
  recommendation}. In \bibinfo{booktitle}{\emph{Proceedings of the ACM SIGKDD
  Conference on Knowledge Discovery \& Data Mining 2021}}.
  \bibinfo{pages}{2084--2092}.
\newblock


\bibitem[Yu et~al\mbox{.}(2021b)]%
        {02yu2021self}
\bibfield{author}{\bibinfo{person}{Junliang Yu}, \bibinfo{person}{Hongzhi Yin},
  \bibinfo{person}{Jundong Li}, \bibinfo{person}{Qinyong Wang},
  \bibinfo{person}{Nguyen Quoc~Viet Hung}, {and} \bibinfo{person}{Xiangliang
  Zhang}.} \bibinfo{year}{2021}\natexlab{b}.
\newblock \showarticletitle{Self-supervised multi-channel hypergraph
  convolutional network for social recommendation}. In
  \bibinfo{booktitle}{\emph{Proceedings of the Web Conference 2021}}.
  \bibinfo{pages}{413--424}.
\newblock


\bibitem[Yu et~al\mbox{.}(2022a)]%
        {15yu2022graph}
\bibfield{author}{\bibinfo{person}{Junliang Yu}, \bibinfo{person}{Hongzhi Yin},
  \bibinfo{person}{Xin Xia}, \bibinfo{person}{Tong Chen},
  \bibinfo{person}{Lizhen Cui}, {and} \bibinfo{person}{Quoc Viet~Hung Nguyen}.}
  \bibinfo{year}{2022}\natexlab{a}.
\newblock \showarticletitle{Are graph augmentations necessary? simple graph
  contrastive learning for recommendation}. In
  \bibinfo{booktitle}{\emph{Proceedings of the International ACM SIGIR
  Conference on Research and Development in Information Retrieval 2022}}.
  \bibinfo{pages}{1294--1303}.
\newblock


\bibitem[Yu et~al\mbox{.}(2022b)]%
        {59yu2022self}
\bibfield{author}{\bibinfo{person}{Junliang Yu}, \bibinfo{person}{Hongzhi Yin},
  \bibinfo{person}{Xin Xia}, \bibinfo{person}{Tong Chen},
  \bibinfo{person}{Jundong Li}, {and} \bibinfo{person}{Zi Huang}.}
  \bibinfo{year}{2022}\natexlab{b}.
\newblock \showarticletitle{Self-supervised learning for recommender systems: A
  survey}.
\newblock \bibinfo{journal}{\emph{arXiv preprint arXiv:2203.15876}}
  (\bibinfo{year}{2022}).
\newblock


\bibitem[Yuan et~al\mbox{.}(2018)]%
        {07yuan2018fbgd}
\bibfield{author}{\bibinfo{person}{Fajie Yuan}, \bibinfo{person}{Xin Xin},
  \bibinfo{person}{Xiangnan He}, \bibinfo{person}{Guibing Guo},
  \bibinfo{person}{Weinan Zhang}, \bibinfo{person}{Chua Tat-Seng}, {and}
  \bibinfo{person}{Joemon~M Jose}.} \bibinfo{year}{2018}\natexlab{}.
\newblock \showarticletitle{fBGD: Learning embeddings from positive unlabeled
  data with BGD}.
\newblock  (\bibinfo{year}{2018}).
\newblock


\bibitem[Zhang et~al\mbox{.}(2021)]%
        {22zhang2021double}
\bibfield{author}{\bibinfo{person}{Junwei Zhang}, \bibinfo{person}{Min Gao},
  \bibinfo{person}{Junliang Yu}, \bibinfo{person}{Lei Guo},
  \bibinfo{person}{Jundong Li}, {and} \bibinfo{person}{Hongzhi Yin}.}
  \bibinfo{year}{2021}\natexlab{}.
\newblock \showarticletitle{Double-scale self-supervised hypergraph learning
  for group recommendation}. In \bibinfo{booktitle}{\emph{Proceedings of the
  ACM International Conference on Information \& Knowledge Management 2021}}.
  \bibinfo{pages}{2557--2567}.
\newblock


\bibitem[Zhang et~al\mbox{.}(2019)]%
        {03zhang2019nonlinear}
\bibfield{author}{\bibinfo{person}{Junwei Zhang}, \bibinfo{person}{Min Gao},
  \bibinfo{person}{Junliang Yu}, \bibinfo{person}{Xinyi Wang},
  \bibinfo{person}{Yuqi Song}, {and} \bibinfo{person}{Qingyu Xiong}.}
  \bibinfo{year}{2019}\natexlab{}.
\newblock \showarticletitle{Nonlinear Transformation for Multiple Auxiliary
  Information in Music Recommendation}. In
  \bibinfo{booktitle}{\emph{Proceedings of International Joint Conference on
  Neural Networks 2019}}. IEEE, \bibinfo{pages}{1--8}.
\newblock


\bibitem[Zhang et~al\mbox{.}(2022)]%
        {44zhang2022neuro}
\bibfield{author}{\bibinfo{person}{Wei Zhang}, \bibinfo{person}{Junbing Yan},
  \bibinfo{person}{Zhuo Wang}, {and} \bibinfo{person}{Jianyong Wang}.}
  \bibinfo{year}{2022}\natexlab{}.
\newblock \showarticletitle{Neuro-Symbolic Interpretable Collaborative
  Filtering for Attribute-based Recommendation}. In
  \bibinfo{booktitle}{\emph{Proceedings of the ACM Web Conference 2022}}.
  \bibinfo{pages}{3229--3238}.
\newblock


\bibitem[Zhao et~al\mbox{.}(2021)]%
        {18zhao2021dataset}
\bibfield{author}{\bibinfo{person}{Bo Zhao}, \bibinfo{person}{Konda~Reddy
  Mopuri}, {and} \bibinfo{person}{Hakan Bilen}.}
  \bibinfo{year}{2021}\natexlab{}.
\newblock \showarticletitle{Dataset Condensation with Gradient Matching.}
\newblock \bibinfo{journal}{\emph{Proceedings of International Conference on
  Learning Representations 2021}} \bibinfo{volume}{1}, \bibinfo{number}{2}
  (\bibinfo{year}{2021}), \bibinfo{pages}{3}.
\newblock


\bibitem[Zhao et~al\mbox{.}(2014)]%
        {21zhao2014leveraging}
\bibfield{author}{\bibinfo{person}{Tong Zhao}, \bibinfo{person}{Julian
  McAuley}, {and} \bibinfo{person}{Irwin King}.}
  \bibinfo{year}{2014}\natexlab{}.
\newblock \showarticletitle{Leveraging social connections to improve
  personalized ranking for collaborative filtering}. In
  \bibinfo{booktitle}{\emph{Proceedings of the ACM international conference on
  conference on information and knowledge management 2014}}.
  \bibinfo{pages}{261--270}.
\newblock


\bibitem[Z{\"u}gner and G{\"u}nnemann(2018)]%
        {46zugner2018adversarial}
\bibfield{author}{\bibinfo{person}{Daniel Z{\"u}gner} {and}
  \bibinfo{person}{Stephan G{\"u}nnemann}.} \bibinfo{year}{2018}\natexlab{}.
\newblock \showarticletitle{Adversarial Attacks on Graph Neural Networks via
  Meta Learning}. In \bibinfo{booktitle}{\emph{Proceedings of International
  Conference on Learning Representations 2018}}.
\newblock


\end{thebibliography}

\clearpage
\appendix

\section{APPENDIX}
\subsection{AU Loss}
\label{wau}
Recent studies \cite{33wang2020understanding} have proved alignment and uniformity of representations are related to prediction performance in representation learning, which is also applicable to the recommendation model \cite{15yu2022graph,17wang2022towards}. The alignment is defined as the expected distance between normalized embeddings of positive pairs, and on the other hand, uniformity is defined as the logarithm of the average pairwise Gaussian potential. AU (Alignment and Uniformity) loss \cite{17wang2022towards} quantified alignment and uniformity in recommendation as follows: 

\begin{equation}
\label{equation11}
\begin{split}
L_{rec}=L_{alignment}+\gamma L_{uniformity},
\end{split}
\end{equation}

\begin{equation}
\label{equation12}
\begin{split}
L_{alignment} = \underset{(u, i, j) \sim P_{\mathcal{D}}}{\mathbb{E}} \vert \vert \widetilde f({z}_{u})-\widetilde f({z}_{i})\vert \vert^{2}, 
\end{split}
\end{equation}

\begin{equation}
\label{equation13}
\begin{split}
L_{uniformity} = (log \underset{(u, u^{'}) \sim P_{U}}{\mathbb{E}} e^{-2\vert \vert \widetilde f({z}_{u})-\widetilde f({z}_{u^{'}}) \vert \vert^{2}}
    \\+log \underset{(i, i^{'}) \sim P_{I}}{\mathbb{E}}e^{-2\vert \vert \widetilde f({z}_{i})-\widetilde f({z}_{i^{'}})\vert \vert^{2}})/2,
\end{split}
\end{equation}
where $P_{U}(\cdot)$ and $P_{I}(\cdot)$ are the distributions of the user set and item set, respectively. $\vert \vert \cdot \vert \vert$ is the $l_{1}$ distance, and $\widetilde{f}(\cdot)$ indicates the $l_{2}$ normalized representations of $f(\cdot)$. The weight $\gamma$ controls the desired degree of uniformity.

\end{document}